\documentclass[letterpaper, english, aps, prl, reprint, twocolumn, superscriptaddress]{revtex4-2}
\usepackage[T1]{fontenc}
\usepackage{color}
\usepackage{amsmath}
\usepackage{amssymb}
\usepackage{graphicx}
\usepackage{multirow}
\usepackage{booktabs}
\usepackage{capt-of}
\usepackage{babel}
\usepackage{calc}
\usepackage{float}
\usepackage{dcolumn}
\usepackage{gensymb}

\makeatletter
\usepackage[colorlinks=true, letterpaper=true, pdfstartview=FitV, linkcolor=blue, citecolor=blue, urlcolor=blue]{hyperref}
\renewcommand{\fnum@figure}{FIG. \thefigure}
\renewcommand{\fnum@table}{TABLE. \thetable}
\makeatother

\begin{document}
\title{Antiferroaxial altermagnetism}
\author{Yichen Liu}
\author{Cheng-Cheng Liu}
\email{ccliu@bit.edu.cn}
\affiliation{Centre for Quantum Physics, Key Laboratory of Advanced Optoelectronic Quantum Architecture and Measurement (MOE), School of Physics, Beijing Institute of Technology, Beijing 100081, China}

\begin{abstract}
The antiferroaxial state is emerging as an important ferroic order in condensed matter systems. Here, we establish antiferroaxial altermagnetism as a broadly prevalent, generic, and microscopically grounded multiferroic mechanism, in which antiferroaxial counter-rotating distortions both induce altermagnetism and enable its deterministic and reversible switching. Within a unified Landau-theory and symmetry framework, we identify a symmetry-allowed trilinear invariant coupling the antiferroaxial order, the N\'{e}el vector, and the altermagnetic order, and derive general symmetry criteria for its occurrence. This coupling locks the induced altermagnetism to the antiferroaxial order, so that reversing the latter reverses the spin splitting and associated time-reversal-odd responses, such as anomalous Hall conductivity. We provide a practical spin group dictionary mapping N\'{e}el-vector representations to the resulting $d$-, $g$-, and $i$-wave antiferroaxial altermagnetism, validate the mechanism with ligand-rotation tight-binding models and first-principles calculations, and identify many candidate materials by screening the MAGNDATA and C2DB databases. Our results elevate antiferroaxiality to a universal ferroic control knob for structurally programmable altermagnetic spintronics.
\end{abstract}
\maketitle
\begin{figure}[t]
    \centering
    \includegraphics[width=0.40\textwidth]{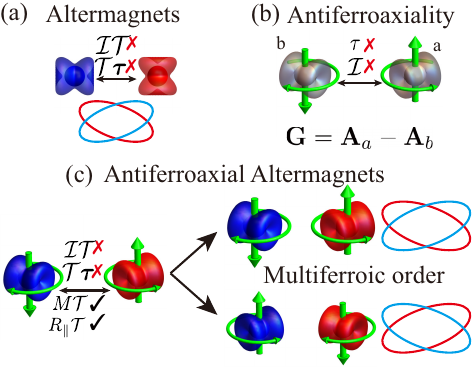}
    \caption{\label{fig:fig1} Antiferroaxial altermagnetism and multiferroic control of the spin splitting. (a) Altermagnet with a spin-splitting isoenergy surface with the red and blue denoting magnetic sublattices; both inversion-time ($\mathcal{IT}$) and translation-time-reversal ($\mathcal{T}\boldsymbol\tau$) symmetries are broken.
  (b) Antiferroaxiality, characterized by counter-rotating sublattices, is defined as $\mathbf{G} = \mathbf{A}_a - \mathbf{A}_b$ with $\mathbf{A}_{a/b}$ the ferroaxial order on the sublattice $a/b$. Antiferroaxial order breaks $\mathcal I$ and $\boldsymbol\tau$ that swap the two sublattices, but preserves those acting within each sublattice.
  (c) In antiferroaxial altermagnets, the coexistence of N\'{e}el and antiferroaxial orders breaks $\mathcal{IT}$ and $\mathcal{T}\boldsymbol\tau$ while preserving mirror-time ($M\mathcal{T}$) and rotation-time ($R_{\parallel}\mathcal{T}$) symmetries, where $\parallel$ denotes the axis parallel to the antiferroaxial order. Antiferroaxial altermagnetism realizes a broadly prevalent and generic multiferroic mechanism enabling antiferroaxiality-controlled altermagnetic spin splitting.}
\end{figure}

\textit{Introduction.---}
Altermagnetism has attracted widespread attention due to its coexistence of nonrelativistic spin-splitting bands and zero net magnetization, thereby combining key features of ferromagnets and antiferromagnets~\cite{smejkalConventionalFerromagnetismAntiferromagnetism2022,
smejkalGiantTunnelingMagnetoresistance2022,
mazinEditorialAltermagnetismANew2022,
wuFermiLiquidInstabilities2007a,
hayamiMomentumDependentSpinSplitting2019c,
yuanGiantMomentumdependentSpin2020b,
maMultifunctionalAntiferromagneticMaterials2021,
liuSpinGroupSymmetryMagnetic2022a,
shaoSpinneutralCurrentsSpintronics2021,
fengAnomalousHallEffect2022a,
baiEfficientSpintoChargeConversion2023,
zhouCrystalThermalTransport2024c,
jinSkyrmionHallEffect2024,
linCoulombDragAltermagnets2025,
laiWaveFlatFermi2025,
liMajoranaCornerModes2023a,
zhuTopologicalSuperconductivityTwodimensional2023b,
zhangFinitemomentumCooperPairing2024b,
ghorashiAltermagneticRoutesMajorana2024,
vijayvargiaAltermagnetsTopologicalOrder2025,
heNonrelativisticSpinMomentumCoupling2023,
panGeneralStackingTheory2024,
liuTwistedMagneticVan2024a,
zhangPredictableGateFieldControl2024,
zhouContrastingLightInducedSpin2025,
zhuDesignAltermagneticModels2025,
chenElectricalSwitchingAltermagnetism2025,
cheEngineeringAltermagneticStates2025,
kaushalAltermagnetismModifiedLieb2025,
leebSpontaneousFormationAltermagnetism2024b,
wangSpinOrbitalAltermagnetism2025a,
caoSymmetryClassificationAlternating2025,
gaoAIacceleratedDiscoveryAltermagnetic2025,
chengOrientationdependentJosephsonEffect2024,
guoValleyPolarizationTwisted2024,
zengBilayerStacking$A$type2024,
ezawaThirdorderFifthorderNonlinear2025,
wangElectricFieldInducedSwitchableTwoDimensional2025,
liFerrovalleyPhysicsStacked2025,
songUnifiedSymmetryClassification2025}.
This unconventional property stems from rotational or mirror symmetries that relate the magnetic sublattices, rather than from translation or inversion symmetries, as illustrated in Fig.~\ref{fig:fig1}(a).
In the limit of vanishing spin-orbit coupling (SOC), the N\'{e}el vector $\mathbf{N}$ of altermagnets transforms as $\Gamma_{\text{AM}}\otimes\Gamma_A^s$, where $\Gamma_{\mathrm{AM}}$ is a nontrivial, inversion-even, one-dimensional (1D) irreducible representation (IR) of the little group at the $\Gamma$ point, and $\Gamma_A^s$ denotes the axial-vector IR of the spin-rotation group. The hallmark altermagnetic nonrelativistic spin splitting is characterized by a magnetic multipole $\mathbf{O}^{(\ell)} \equiv \int d^3 r [r_{\mu_1}r_{\mu_2}\cdots r_{\mu_\ell}]\mathbf{m}(\mathbf{r})$ with $\mathbf m(\mathbf r)$ the magnetization density and $r_\mu$ the Cartesian components~\cite{mcclartyLandauTheoryAltermagnetism2024,schiffCollinearAltermagnetsTheir2025}. The brackets $[\cdots]$ indicate the symmetric-traceless projection over $\mu_1\cdots\mu_\ell$, so that $\mathbf O^{(\ell)}$ transforms as $\Gamma_\ell\otimes\Gamma_A^s$, with $\Gamma_\ell$ the rank-$\ell$ spatial IR.
Recent studies have reported that these altermagnetic multipoles can couple to classical multiferroic orders~\cite{smejkalAltermagneticMultiferroicsAltermagnetoelectric2024,sunAltermagnetismInducedSliding2024,duanAntiferroelectricAltermagnetsAntiferroelectricity2025,guFerroelectricSwitchableAltermagnetism2025,zhuSlidingFerroelectricControl2025,huangSpinInversionEnforced2025a,dingFerroelasticallyTunableAltermagnets2025,pengFerroelasticAltermagnetism2025a} , such as the ferroelectric and ferroelastic ones, suggesting a path toward altermagnetic multiferroicity.

In addition to conventional ferromagnetic, ferroelectric, and ferroelastic orders, there is a new type of significant ferroic order, the ferroaxial order~\cite{hayashidaObservationFerrochiralTransition2021,hlinkaEightTypesSymmetrically2014,johnsonGiantImproperFerroelectricity2012,hlinkaSymmetryGuideFerroaxial2016,singhFerroaxialDensityWave2025,zengPhotoinducedNonvolatileRewritable2025,heOpticalControlFerroaxial2024,day-robertsPiezoresistivityFingerprintFerroaxial2025,jinObservationFerrorotationalOrder2020}. Ferroaxiality is characterized by the rotational distortion of structural units, represented by the axial vector $\mathbf{A}$~\cite{1986JETP}, whose magnitude is parameterized by the rotation angle $\theta$. The ferroaxial order transforms according to the axial-vector IR $\Gamma_A$ and is invariant under inversion and time reversal symmetry. Analogously, the antiferroaxial order is characterized by structural units exhibiting counter-rotating distortions of equal magnitude but opposite directions along a shared axis~\cite{hayashidaObservationFerrochiralTransition2021}, as illustrated in Fig.~\ref{fig:fig1}(b). This state is characterized by a staggered order parameter $\mathbf{G} = \mathbf{A}_a - \mathbf{A}_b$, where $\mathbf{A}_a$ and $\mathbf{A}_b$ denote the local ferroaxial vectors of sublattices $a$ and $b$, respectively. Antiferroaxial order commonly occurs in many systems, such as perovskites and transition metal trifluorides~\cite{autretStructuralInvestigationCa2MnO42004,knightNuclearMagneticStructures2020,leeWeakferromagnetismCoF3FeF32018,mattssonDensityFunctionalTheory2019,sodequistTwodimensionalAltermagnetsHigh2024c}. Unlike the ferroelectric order, the ferroaxial and antiferroaxial orders do not require inversion symmetry breaking and can couple to physical quantities such as strain, optical fields, and piezoresistance~\cite{zengPhotoinducedNonvolatileRewritable2025,heOpticalControlFerroaxial2024,jinObservationFerrorotationalOrder2020,day-robertsPiezoresistivityFingerprintFerroaxial2025}.

In this Letter, we propose antiferroaxial altermagnetism, a multiferroic state in which altermagnetism is generated and reversibly switched by the antiferroaxial order.
Using symmetry analysis and Landau theory, we demonstrate a symmetry-allowed trilinear coupling among the antiferroaxial order $\mathbf{G}$, the N\'{e}el order $\mathbf{N}$, and magnetic multipoles $\mathbf{O}^{(\ell)}$, and establish the general symmetry criteria governing this interaction. This coupling dictates that reversing the antiferroaxial order inverts the spin splitting, enabling multiferroic control. Furthermore, Table~\ref{table:TB1} classifies the altermagnetic spin point groups and their associated $\Gamma_{\mathrm{AM}}$, serving as a practical guide for determining the spin splitting type of antiferroaxial altermagnetism. We illustrate antiferroaxial altermagnetism via tight-binding models incorporating ligand rotations, confirming its applicability to generate and control $d$-, $g$-, and $i$-wave altermagnetism.
Using first-principles calculations combined with screening the MAGNDATA~\cite{gallego2016magndata} and C2DB~\cite{gjerdingRecentProgressComputational2021,haastrupComputational2DMaterials2018} databases, we identify many candidate antiferroaxial altermagnets (Table~\ref{tab:table2}).
Taking FeF$_3$ as an example, we demonstrate that reversing the antiferroaxial order not only switches the spin splitting but also inverts the time-reversal-odd responses, such as the anomalous Hall conductivity (AHC), establishing a novel mechanism for structural control of spintronics.

\textit{Landau theory.---} In Landau theory, an antiferroaxial order $\mathbf{G}$ transforms as $\Gamma_G(\mathbf{q})\otimes\Gamma_A$ in the parent space group $\mathcal{G}_0$, where $\Gamma_G(\mathbf{q})$ is the IR of the little group at the propagation vector $\mathbf{q}$ that encodes the staggered sublattice pattern and $\Gamma_A$ denotes the axial-vector representation. Similarly, in the zero SOC limit, the N\'{e}el vector $\mathbf{N}$ transforms as $\Gamma_N(\mathbf{q}') \otimes \Gamma_A^{s}$, where $\Gamma_N(\mathbf{q}')$ is the IR of the little group at $\mathbf{q}'$. A symmetry-allowed trilinear coupling among $\mathbf{G}$, $\mathbf{N}$, and $\mathbf{O}^{(\ell)}$ requires that the product of their representations contains the identity representation. Since the spin-rotation sector satisfies $\Gamma_1^s \subset \Gamma_A^s \otimes \Gamma_A^s$, where $\Gamma_1^s$ is the identity representation in spin-rotation space, the invariance condition reduces to the spatial sector only,
\begin{equation}
\Gamma_1 \subset\Gamma_G(\mathbf{q}) \otimes \Gamma_A\otimes \Gamma_N(\mathbf{q}') \otimes \Gamma_{\ell}.
\label{Eq:AM_quest}
\end{equation}
This implies that the magnetic multipole $\mathbf{O}^{(\ell)}$ can be generally induced by the coexistence of the antiferroaxial order and the N\'{e}el vector.
Moreover, translational invariance of the coupling enforces momentum conservation, $\mathbf{q}' = -\mathbf{q}$. Equation~(\ref{Eq:AM_quest}) identifies all symmetry-allowed $\mathbf{O}^{(\ell)}$; specifically, the lowest rank $\ell$ dictates the type ($d$-, $g$-, or $i$-wave) of the induced altermagnetism. However, the spin-splitting type (planar or bulk) is established only upon the condensation of $\mathbf{G}$. Once $\mathbf{G}$ condenses and reduces the symmetry to the isotropy subgroup $\mathcal{G}$, the N\'{e}el vector's representation in $\mathcal{G}$ is obtained by subducing the parent representation, i.e., $\Gamma_N(\mathbf{q}') \downarrow \mathcal{G}$ (see Supplemental Material (SM)~\cite{SM}). Equation~(\ref{Eq:AM_quest}) can be stated equivalently as follows: a system is an antiferroaxial altermagnet if and only if, after the antiferroaxial order $\mathbf{G}$ condenses and lowers the symmetry to the isotropy subgroup $\mathcal G$, the subduced representation of the N\'{e}el vector, $\Gamma_N(\mathbf q')\downarrow \mathcal G$, contains a 1D IR that matches one of the $\Gamma_{\mathrm{AM}}$ listed in Table~\ref{table:TB1}, which maps each $\Gamma_{\mathrm{AM}}$ to its corresponding splitting type and collinear spin point group.

The coupling dictates that reversing $\mathbf{G}$ at fixed $\mathbf{N}$ necessarily enforces a sign reversal of the magnetic multipole $\mathbf{O}^{(\ell)}$ and the associated spin splitting, thereby establishing the multiferroic mechanism in which a nonmagnetic antiferroaxial order deterministically controls magnetic responses. Unlike ferroelectric orders, the antiferroaxial order does not necessarily break inversion symmetry, allowing this multiferroic control to operate even in non-polar, centrosymmetric systems.

\begin{figure}[t]
    \centering
    \includegraphics[width=0.49\textwidth]{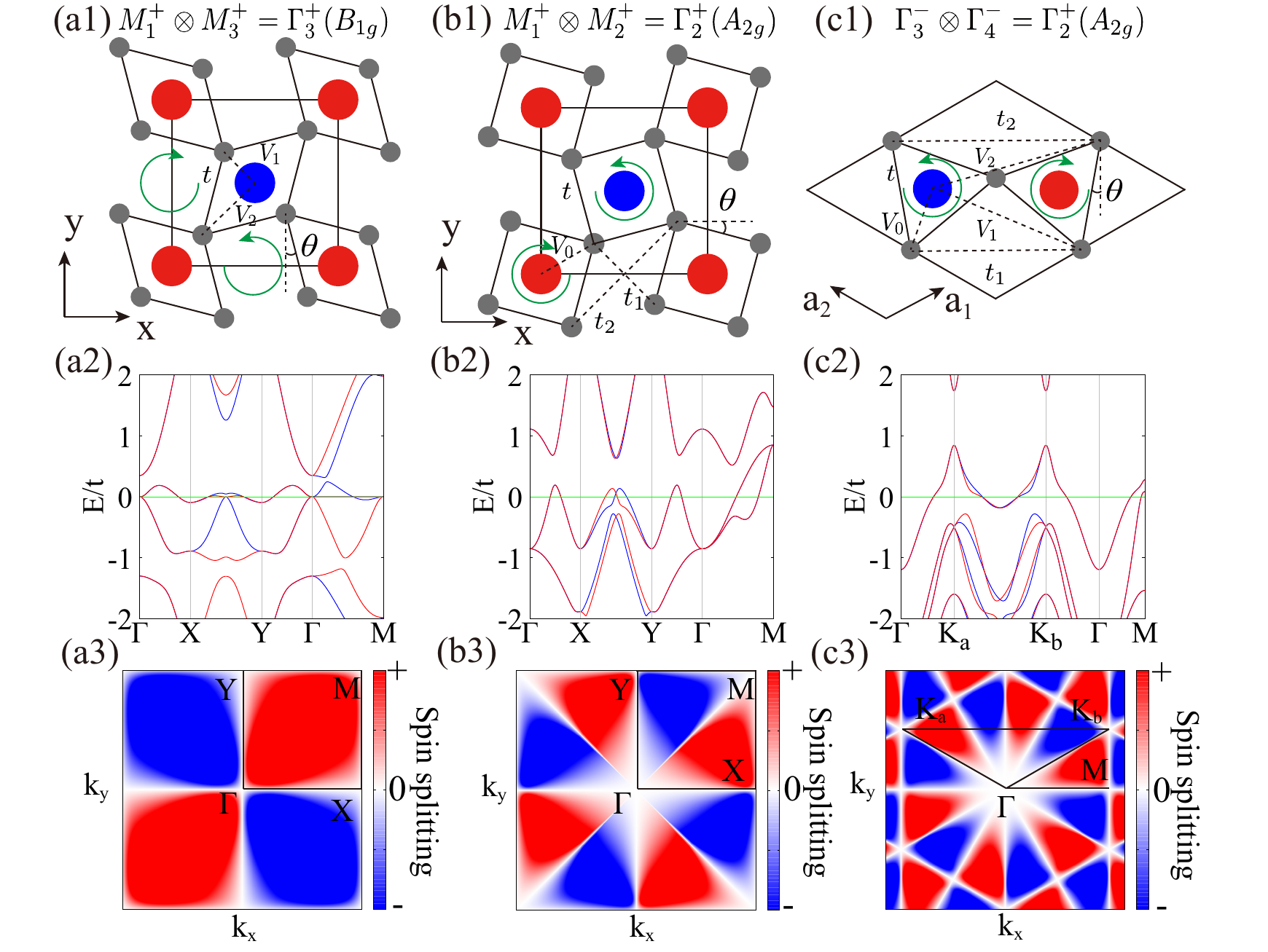}
    \caption{\label{fig:fig2} Antiferroaxial altermagnetism with distinct spin splitting types.
Red and blue spheres denote spin-up and spin-down atoms, gray spheres represent ligands, and green arrows indicate rotational distortions.
(a)-(c) Illustration of the mechanism (top), band structures (middle), and spin splittings (bottom).
(a) $d$-wave altermagnetism from parent group $P4/mmm$: $N\sim M_1^+$ couples to $G_z\sim M_3^+$ via a $\mathbf{O}^{(2)}$ magnetic multipole in the $\Gamma_3^+(B_{2g})$ channel (Table~S1 in SM~\cite{SM}). Upon symmetry lowering to $P4/mbm$, the N\'{e}el vector transforms as $\Gamma_4^+(B_{2g})$, giving a $d_{xy}$-wave spin splitting in point group $4/mmm$ (Table~\ref{table:TB1}).
(b) $g$-wave altermagnetism from parent group $P4/mmm$: $N\sim M_1^+$ couples to $G_z\sim M_2^+$ via the $\Gamma_2^+(A_{2g})$ magnetic multipole channel. In $P4/mbm$, the altermagnetic N\'{e}el vector is $\Gamma_2^+(A_{2g})$, corresponding to a planar $g$-wave spin splitting in $4/mmm$ (Table~\ref{table:TB1}).
(c) $i$-wave altermagnetism from parent group $P6/mmm$: $N\sim\Gamma_3^-$ couples to $G_z\sim \Gamma_4^-$ via the $\Gamma_2^+(A_{2g})$ magnetic multipole channel. The symmetry lowering to $P\bar{6}m2$ gives an altermagnetic N\'{e}el vector $\Gamma_2(A_2^\prime)$, corresponding to a planar $i$-wave spin splitting in $\bar{6}m2$ (Table~\ref{table:TB1}).
  The common model parameters are $J=t$ and $t_M=0.5t$, with specific values:
    (a) $\theta=20^\circ$, $V_1=1.69t$, $V_2=0.37t$;
    (b) $\theta=20^\circ$, $V_0=1.2t$, $t_1=0.7t$, $t_2=0.15t$;
    (c) $\theta=30^\circ$, $V_0=1.2t$, $V_1=0.875t$, $V_2=0.5t$, $t_1=0.6t$, $t_2=0.15t$.}
\end{figure}

\textit{Model.---} To clarify the microscopic mechanism, we perform the explicit symmetry analysis and construct tight-binding models that incorporate the ligand degrees of freedom. This allows us to directly map the antiferroaxial order onto the electronic structure. The model is defined on a lattice with magnetic atoms on sublattices $a$ and $b$, bridged by nonmagnetic ligands whose rotational distortion encodes the antiferroaxial order [see Fig.~\ref{fig:fig2}]. The corresponding Hamiltonian reads (see SM~\cite{SM})
\begin{equation}
\begin{split}
\mathcal{H} =
&t_M\sum_{\langle mn\rangle} d_m^\dagger d_n
+ \sum_{mi}\left[ V_{mi}(\theta)\, d_m^\dagger c_i + \mathrm{h.c.} \right] \\
&+ \sum_{ij} t_{ij}(\theta)\, c_i^\dagger c_j
+ J\sum_m \eta_m\, d_m^\dagger \sigma_z d_m.
\end{split}
\label{eq:hamiltonian}
\end{equation}
Here, $d_m^\dagger$ ($c_i^\dagger$) denotes the creation operator for an electron on magnetic site $m$ (ligand site $i$).
The first term represents the hopping $t_M$ between nearest-neighbor magnetic sites.
The second and third terms describe the magnetic-ligand hybridization $V_{mi}(\theta)$ and the ligand-ligand hopping $t_{ij}(\theta)$, respectively.
Crucially, the hopping amplitudes such as $t_{1/2}$ and $V_{1/2}$ vary with the rotation angle $\theta$ due to changes in corresponding bond lengths [Figs.~\ref{fig:fig2}(a1)-(c1)]. This structural dependence is incorporated via Harrison's scaling law $t(r) \propto r^{-2}$~\cite{PhysRevB.20.2420}, where the explicit forms of $V_{mi}(\theta)$ and $t_{ij}(\theta)$ are given in SM~\cite{SM}.
The sums over $mi$ and $ij$ run over the symmetry-inequivalent pathways ($V_{1/2}$ and $t_{1/2}$) illustrated in Fig.~\ref{fig:fig2}.
The final term introduces the local antiferromagnetic exchange $J$, where $\sigma_z$ is the spin operator and $\eta_m = \pm 1$ encodes the staggered N\'{e}el order.

Figures~\ref{fig:fig2}(a1)-(a3) show the geometry, band structure, and spin splitting of the $d$-wave antiferroaxial altermagnetic model. The antiferroaxial order is parameterized by the ligand rotation angle $\theta$, as illustrated in Fig.~\ref{fig:fig2}(a1).
The parent phase belongs to space group $P4/mmm$. As a representative scenario, we consider the transitions at the $M$ point, $\mathbf{q}=(\pi,\pi,0)$, for which $\mathbf{N}\sim M_1^+$ and $\mathbf{G}\sim M_4^+\otimes(\Gamma_2^+\oplus\Gamma_5^+)$. For the case in which $\mathbf G$ is aligned along the z direction, $G_z\sim M_4^+\otimes\Gamma_2^+=M_3^+$. The bilinear combination thus transforms as $\mathbf{N}G_z\sim M_1^+\otimes M_3^+= \Gamma_3^+(B_{1g})$, where $B_{1g}$ follows the Mulliken label, while $M_i^+$ and $\Gamma_i^+$ follow the Bradley-Cracknell convention~\cite{BC_form}.
In the $P4/mmm$ group, the rank-2 multipole representation decomposes as $\mathbf{O}^{(2)}\sim A_{1g}\oplus B_{1g}\oplus B_{2g}\oplus E_g$ (see Table~S1 in SM~\cite{SM}), so the representation constraint in Eq.~(\ref{Eq:AM_quest}) is satisfied and a symmetry-allowed trilinear invariant exists.
When $G_z$ condenses, the system symmetry is reduced to $P4/mbm$ [Fig.~\ref{fig:fig2}(a1)], which has crystallographic point group $4/mmm$. The propagation vector $\mathbf{q}$ is folded to the Brillouin-zone center, and the N\'{e}el vector transforms as $\mathbf{N}\sim B_{2g}$ (see details in SM~\cite{SM}).
Table~\ref{table:TB1} identifies this IR in point group $4/mmm$ corresponding to a $d$-wave spin splitting, with the associated altermagnetic spin point group $^{\bar{1}}$4/m$^{\bar{1}}$mm. We use Litvin notation for spin point groups here~\cite{litvinSpinPointGroups1977}. The numerical results of our generic TB model with $\theta=20^\circ$ also show the same $d$-wave spin splitting. Reversing $\theta$, i.e., reversing $\mathbf{G}$, is symmetry-equivalent to the operation $[E || C_{4z}]$, thereby reversing the sign of the spin splitting (see details in SM~\cite{SM}). Here $[E|| C_{4z}]$ is a spin-group operation, where $E$ denotes the identity operation in spin space and $C_{4z}$ denotes a four-fold rotation in real space.

For the $g$-wave case, the parent group is also $P4/mmm$. In contrast to Fig.~\ref{fig:fig2}(a), here the antiferroaxial order and the N\'{e}el order share the same spatial pattern with $\mathbf{N}\sim M_1^+$ and $G_z\sim M_1^+\otimes\Gamma_2^+=M_2^+$. The corresponding bilinear combination transforms as
$\mathbf{N}G_z\sim M_1^+ \otimes M_2^+= \Gamma_2^+(A_{2g})$,
which cannot couple to $\mathbf{O}^{(2)}$ but can couple to the $A_{2g}$ channel of the $\mathbf{O}^{(4)}$ (see Table~S1 in SM~\cite{SM}). After $G_z$ condenses, the symmetry is reduced to $P4/mbm$ [Fig.~\ref{fig:fig2}(b1)] and the N\'{e}el vector transforms as $\mathbf{N}\sim A_{2g}$, corresponding to planar $g$-wave altermagnetism (Table~\ref{table:TB1}). The geometry, band structure, and spin splitting are shown in Figs.~\ref{fig:fig2}(b1)-(b3), with parameters $\theta=20^\circ$.
Reversing $\mathbf{G}$ is symmetry-equivalent to $[C_2 ||\boldsymbol\tau]$ ($\boldsymbol\tau=(1/2,1/2,0)$), indicating a spin splitting reversal (see details in SM~\cite{SM}).

Figures~\ref{fig:fig2}(c1)-(c3) show the $i$-wave antiferroaxial altermagnetic model with parent group $P6/mmm$. Both antiferromagnetic and antiferroaxial transitions occur at the $\Gamma$ point, transforming as $\mathbf{N}\sim\Gamma_3^-$ and $G_z\sim\Gamma_3^- \otimes \Gamma_2^+ = \Gamma_4^-$, respectively.
Their coupling yields $\mathbf{N}G_z\sim\Gamma_3^- \otimes \Gamma_4^-= \Gamma_2^+(A_{2g})$,
which can couple to $\mathbf{O}^{(6)} \sim 2A_{1g} \oplus A_{2g} \oplus B_{1g} \oplus B_{2g} \oplus 2E_{1g} \oplus 2E_{2g}$ (see Table~S1 in SM~\cite{SM}). After $G_z$ condenses, the space group reduces to $P\bar{6}m2$, and the N\'{e}el vector transforms as $A_{2}^\prime$, which identifies the system as a planar $i$-wave altermagnet (Table~\ref{table:TB1}), as shown in Figs.~\ref{fig:fig2} (c2) and (c3) with $\theta=30^\circ$.
Reversing the antiferroaxial order is symmetry-equivalent to $[E || m_{110}]$, which also reverses the spin splitting (see details in SM~\cite{SM}).
\begin{figure}[t]
    \centering
    \includegraphics[width=0.48\textwidth]{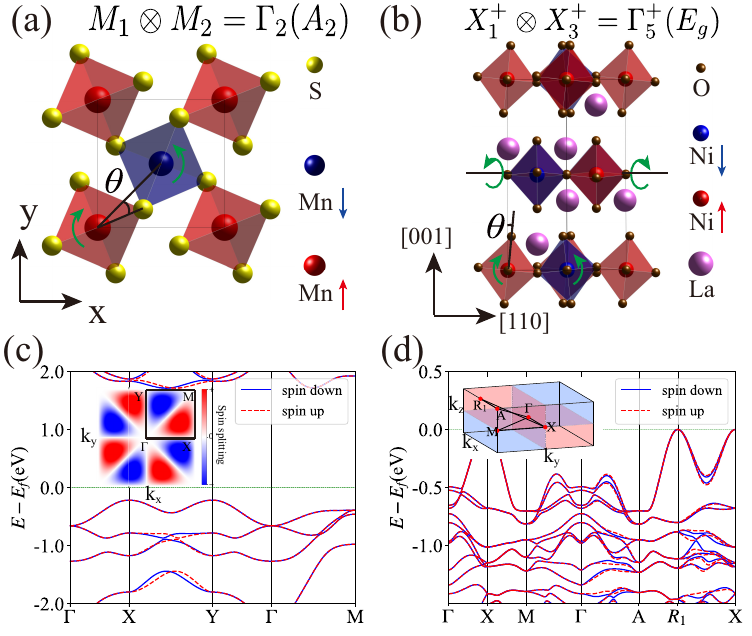}
    \caption{\label{fig:fig3} Antiferroaxial altermagnets. The blue/red spheres and green arrows denote opposite spins and antiferroaxial orders, respectively.
(a, c) Monolayer MnS$_2$ (parent group $P\bar{4}m2$), where N\'{e}el ($M_1$) and antiferroaxial ($M_2$) orders couple with a $\mathbf{O}^{(4)}$ magnetic multipole in the $\Gamma_2(A_2)$ channel (Table~S1 in SM~\cite{SM}). The symmetry lowers to space group $P\bar{4}2_1m$ with N\'{e}el vector $\Gamma_2(A_2)$, corresponding to planar $g$-wave spin splitting (Table~\ref{table:TB1}).
(b, d) Bulk La$_2$NiO$_4$ (parent group $I4/mmm$), where N\'{e}el ($X_1^+$) and antiferroaxial ($X_3^+$) orders couple with the $\Gamma_5^+(E_{g})$ magnetic multipole channel. The symmetry lowers to space group $Pccn$ with N\'{e}el vector $\Gamma_4^+(B_{3g})$, corresponding to $d_{yz}$-wave splitting (Table~\ref{table:TB1}). The insets in the band structures (c, d) illustrate the spin splitting within the Brillouin zone.}
\end{figure}

\textit{Antiferroaxial altermagnets.---} To validate our theoretical framework and demonstrate the ubiquity of the proposed mechanism, we performed a comprehensive screening across the MAGNDATA~\cite{gallego2016magndata,gallegoMAGNDATADatabaseMagnetic2016} and C2DB~\cite{gjerdingRecentProgressComputational2021,haastrupComputational2DMaterials2018} databases. Guided by the above theoretical analysis, we identified a series of candidate materials hosting antiferroaxial altermagnetism, as seen in Table~\ref{tab:table2}. In the following, we present two representative examples: the two-dimensional (2D) $g$-wave altermagnet MnS$_2$~\cite{wangPentagonal2DAltermagnetsb} and the three-dimensional (3D) experimentally synthesized $d$-wave altermagnet La$_2$NiO$_4$~\cite{rodriguez-carvajalNeutronDiffractionStudy1991}, confirming their antiferroaxial altermagnetism via density functional theory (DFT) calculations.

Monolayer MnS$_2$~\cite{wangPentagonal2DAltermagnetsb} is built from Mn-S$_4$ tetrahedra possessing $S_4$ symmetry [Fig.~\ref{fig:fig3}(a)]. Its parent phase belongs to the space group $P\bar{4}m2$. The DFT calculations~\cite{wangPentagonal2DAltermagnetsb} reveal that both antiferroaxial and antiferromagnetic instabilities emerge at the M point ($\mathbf{q}=(\pi,\pi,0)$), leading to an ordered state where the Mn-S tetrahedra exhibit a substantial rotational distortion of $\theta \approx 23.4^\circ$.
Symmetry analysis indicates that the antiferroaxial and antiferromagnetic order parameters transform according to the irreducible representations $G_z\sim M_2$ and $\mathbf{N}\sim M_1$, respectively. Consequently, $G_z\mathbf{N}\sim M_2 \otimes M_1=\Gamma_2(A_2)$ can couple to the $A_2$ channel of the $\mathbf{O}^{(4)}$ (see Table~S1 in SM~\cite{SM}). After $G_z$ condenses, the isotropy subgroup is $P\bar{4}2_1m$ with $\mathbf{N}\sim\Gamma_2(A_2)$, indicating a $g$-wave spin splitting (Table~\ref{table:TB1}), which is corroborated by the DFT results shown in Fig.~\ref{fig:fig3}(c).

The 3D Ruddlesden-Popper compound La$_2$NiO$_4$~\cite{rodriguez-carvajalNeutronDiffractionStudy1991} (parent space group $I4/mmm$) consists of corner-sharing Ni-O$_6$ octahedra [Fig.~\ref{fig:fig3}(b)]. Symmetry-mode analysis indicates concomitant antiferroaxial and antiferromagnetic instabilities at the $X$ point~\cite{hatchCompleteListingOrder2001,stokes2002copl} ($\mathbf{q}=(\pi,\pi,0)$), yielding an antiferroaxial rotation pattern with $\theta \approx 6.3^\circ$ together with antiferromagnetic order, i.e., $G_{110}\sim X_3^+$ and $\mathbf{N}\sim X_1^+$. Their bilinear combination transforms as
$G_{110}\mathbf{N}\sim X_3^+\otimes X_1^+=\Gamma_5^+(E_g)$,
which couples to the $E_g$ channel of the magnetic multipole $\mathbf{O}^{(2)}$, implying an altermagnetic state.
Upon condensation, $G_{110}$ lowers the crystallographic symmetry to $I4_2/ncm$, and the $\mathbf{N}$ further reduces it to space group $Pccn$ with point group $mmm$. In this state, the N\'{e}el order transforms as $\mathbf{N}\sim \Gamma_4^+(B_{3g})$, corresponding to spin point group $^{\bar{1}}mm^{\bar{1}}m$ with $d$-wave altermagnetism (Table~\ref{table:TB1}), confirmed by the band structure calculations [Fig.~\ref{fig:fig3}(d)].
\begin{table}[t]
\caption{\label{tab:table2} The candidate antiferroaxial altermagnets and the corresponding altermagnetic types.}
\footnotesize
\begin{ruledtabular}
\begin{tabular}{ccp{5cm}}
spin splitting & antiferroaxial altermagnets \\
\colrule
$d$-wave & La$_2$NiO$_4$ \\
planar $g$-wave & MnA$_2$ (A=S,C), KMnF$_3$,Ca$_2$MnO$_4$\\
planar $i$-wave & FeBr$_3$\\
bulk $g$-wave & MF$_3$ (M=Fe,Co,Ni), LaCrO$_3$\\
\end{tabular}
\end{ruledtabular}
\end{table}

\begin{figure}[t]
    \centering
    \includegraphics[width=0.48\textwidth]{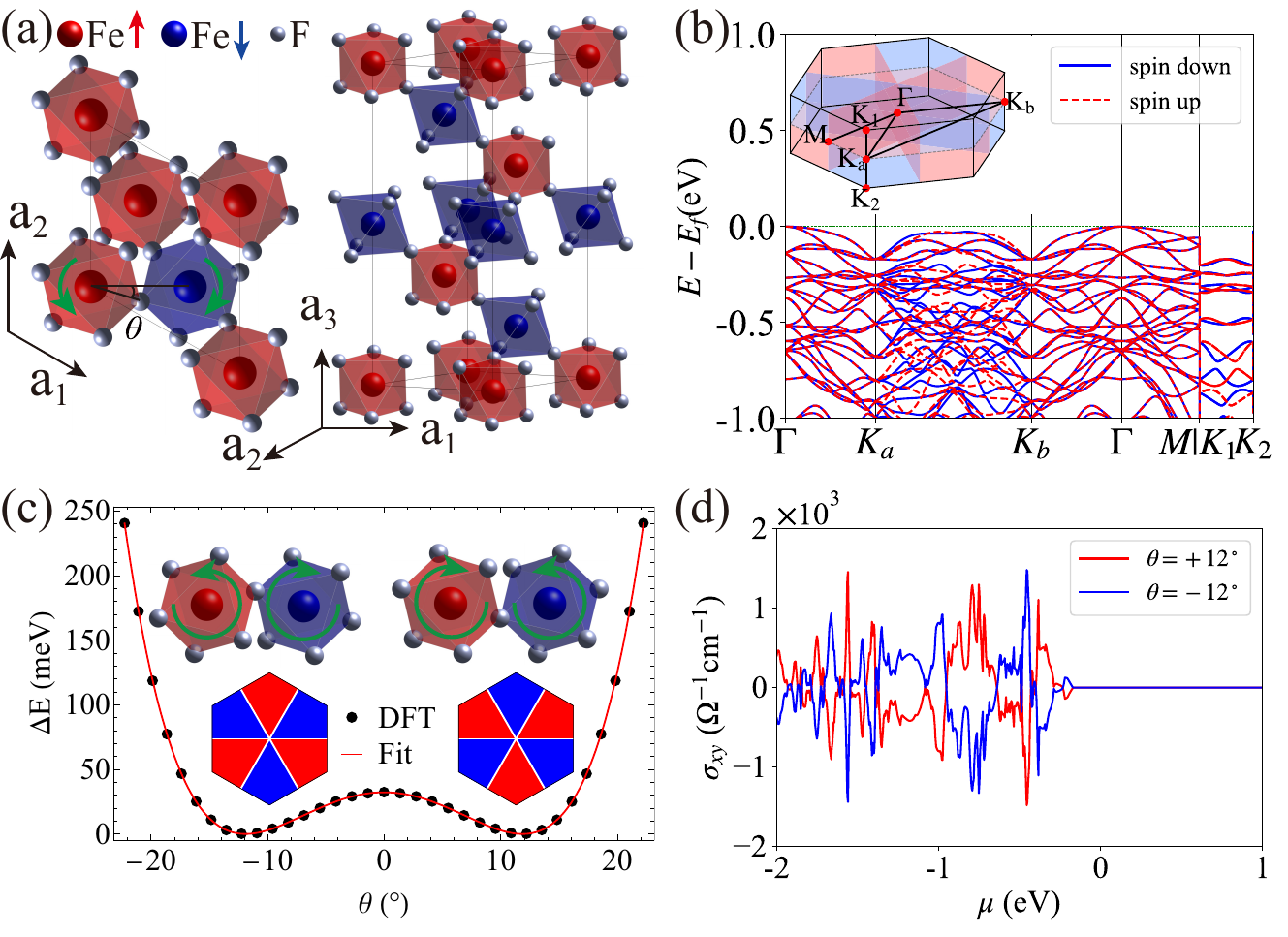}
    \caption{\label{fig:fig4} Multiferroicity with antiferroaxial switchable spin splitting and anomalous Hall conductivity in altermagnets.
(a) Top and side views of the crystal structure of FeF$_3$. Red and blue spheres denote Fe atoms with opposite magnetic moments, while gray spheres represent F atoms. The green curved arrows indicate the antiferroaxial rotation of the Fe-F$_6$ octahedra, parameterized by the angle $\theta$.
(b) Band structure for the lowest-energy configuration at $\theta = 12^\circ$. The inset shows the 3D Brillouin zone with the bulk $g$-wave spin splitting.
(c) Total energy as a function of the rotation angle $\theta$, revealing a double-well potential with ground-state minima at $\theta \approx \pm 12^\circ$. The insets illustrate that reversing the structural rotation direction (top) switches the sign of the spin splitting (bottom). The effective Landau free energy is well fit to the model in Eq.~\ref{Eq:eff_Landau}, yielding $\mathcal{F}_\mathrm{eff}(\theta) = 32.1 - 0.446\theta^2 + 1.45 \times 10^{-3} \theta^4$, where $\theta$ parameterizes the structural distortion $G$ (in degrees) and $\mathcal{F}_\mathrm{eff}$ is in meV per unit cell.
(d) Anomalous Hall conductivity, calculated as a function of chemical potential at the equilibrium angles $\theta = \pm 12^\circ$, is switchable by antiferroaxial order.}
\end{figure}

\textit{Antiferroaxial Multiferroicity.---}
We formulate the Landau theory in terms of the Landau free energy to demonstrate how the antiferroaxial order switches the spin splitting and the associated physical properties, such as AHC, in altermagnets. As an example, we consider the antiferroaxial multiferroicity in the bulk $g$-wave altermagnet FeF$_3$~\cite{leeWeakferromagnetismCoF3FeF32018}. FeF$_3$ is composed of corner-sharing FeF$_6$ octahedra [Fig.~\ref{fig:fig4}(a)]. Starting from the parent space group $Pm\bar{3}m$, it exhibits antiferromagnetic and antiferroaxial instabilities at the $R$ point ($\mathbf{q}=(\pi,\pi,\pi)$). Symmetry analysis gives $\mathbf{G}\sim R_4^+$ and $\mathbf{N}\sim R_1^+$, indicating $\mathbf{G}\mathbf{N}\sim R_{4}^+\otimes R_{1}^+= \Gamma_{4}^+(T_{1g})$. Since the rank-4 magnetic multipole $\mathbf{O}^{(4)}\sim A_{1g}\oplus E_g\oplus T_{1g}\oplus T_{2g}$ contains a $T_{1g}$ spatial channel (see Table~S1 in SM~\cite{SM}), the representation constraint in Eq.~(\ref{Eq:AM_quest}) is satisfied, implying a symmetry-allowed trilinear coupling and a $g$-wave altermagnetism whose planar or bulk character is determined by the orientation of $\mathbf{G}$.

Based on Landau theory, the free energy can be written as (see the SM~\cite{SM}):
\begin{equation}
\begin{split}
\mathcal{F}=&\alpha_N\mathbf{N}^2+\beta_N(\mathbf{N}^2)^2+\alpha_G \mathbf{G}^2+\beta_{G1}(\mathbf{G}^2)^2\\
&+\beta_{G2}(G_x^2 G_y^2+G_x^2 G_z^2+G_y^2 G_z^2)+\mathcal{F}_{\varepsilon}+\mathcal{F}_{\varepsilon G}\\
&+\mathcal{F}_{O}+g\mathbf{N}\cdot(G_x\mathbf{O}_{1}+G_y\mathbf{O}_{2}+G_z\mathbf{O}_{3}),
\end{split}
\end{equation}
where $\mathbf{O}_1=\mathbf{O}_{yz(y^2-z^2)}^{(4)}$, $\mathbf{O}_2=\mathbf{O}_{xz(x^2-z^2)}^{(4)}$, and $\mathbf{O}_3=\mathbf{O}_{xy(x^2-y^2)}^{(4)}$, with subscripts denoting their spatial symmetries. Here $\mathbf{N}^2\equiv\mathbf{N}\cdot\mathbf{N}$ and $\mathbf{G}^2\equiv\mathbf{G}\cdot\mathbf{G}$. The remaining terms $\mathcal{F}_{O}$, $\mathcal{F}_{\varepsilon}$, and $\mathcal{F}_{\varepsilon G}$ denote the magnetic multipole, elastic, and strain-antiferroaxial coupling contributions, respectively. Eliminating the secondary order parameters by their minimization conditions and substituting the equilibrium values back into $\mathcal{F}$, we obtain an effective Landau free energy for $\mathbf{G}$,
\begin{equation}
\mathcal{F}_\text{eff}=\alpha_\text{eff}\mathbf{G}^2+\beta_{1,\text{eff}}(\mathbf{G}^2)^2+\beta_{2,\text{eff}}(G_x^2 G_y^2+G_x^2 G_z^2+G_y^2 G_z^2),
\label{Eq:eff_Landau}
\end{equation}
where the specific expressions of the effective renormalized coefficients $\alpha_\text{eff}$, $\beta_{1,\text{eff}}$ and $\beta_{2,\text{eff}}$ are given in the SM~\cite{SM}. The sign of the anisotropy parameter $\beta_{2,\text{eff}}$ determines the ground state orientation of the antiferroaxial order. When $\beta_{2,\text{eff}} < 0$, the antiferroaxial vector aligns along the $(111)$ direction ($G_x=G_y=G_z$), stabilizing the bulk $g$-wave spin splitting described by $\mathbf{O}^{(4)} \propto (x+y+z)(x-y)(y-z)(z-x)\mathbf{m}(\mathbf{r})$, which is observed in FeF$_3$ [Fig.~\ref{fig:fig4}(b)], NiF$_3$, CoF$_3$, and LaCrO$_3$. Conversely, $\beta_{2,\text{eff}}>0$ favors alignment along a principal axis (e.g., $x, y,$ or $z$), leading to a planar $g$-wave splitting, as observed in KMnF$_3$ and Ca$_2$MnO$_4$~\cite{SM}.

The magnitude of the antiferroaxial order is quantified by the rotation angle $\theta$ of the Fe-F$_6$ octahedra. The DFT calculations reveal a double-well potential for the total energy as a function of $\theta$ [Fig.~\ref{fig:fig4}(c)], with minima located at $\theta=\pm 12^\circ$. The energy barrier is approximately $\Delta E\approx 30$ meV. Although the N\'{e}el vector remains fixed, the two antiferroaxial states ($\pm \theta$) are related by the symmetry operation $[C_2||\boldsymbol{\tau}]$ (where $\boldsymbol{\tau}=(0,0,0.5)$)~\cite{SM}. Consequently, reversing the antiferroaxial order necessitates a reversal of the spin splitting, establishing FeF$_3$ as an antiferroaxial multiferroic altermagnet. Furthermore, in the consideration of the SOC, the two antiferroaxial states are related by the $\mathcal{T}\boldsymbol{\tau}$ symmetry. Since the AHC is odd under time-reversal, this symmetry relation dictates that reversing the antiferroaxial order can simultaneously switch the sign of the AHC, and more generally of any time-reversal-odd response, which is confirmed by the calculations shown in Fig.~\ref{fig:fig4}(d).

\textit{Discussion.---} In this Letter, we demonstrate a novel class of multiferroic altermagnetism, termed antiferroaxial altermagnetism, in which altermagnetism is coupled with and tunable by antiferroaxiality. Based on symmetry analysis and Landau theory, we elucidate the general symmetry criteria governing the trilinear coupling between antiferroaxial, antiferromagnetic, and altermagnetic orders. We substantiate this theory by constructing general ligand-rotation tight-binding models for $d$-, $g$-, and $i$-wave antiferroaxial altermagnetism, confirming the coupling between antiferroaxiality and altermagnetism. By screening the MAGNDATA and C2DB databases combined with \textit{ab initio} calculations, we identified a series of material candidates, exemplifying our theory with the 2D $g$-wave altermagnet MnS$_2$ and the 3D $d$-wave altermagnet La$_2$NiO$_4$. Furthermore, using FeF$_3$ as a prototype, we quantified a switchable energy barrier of 30 meV, demonstrating that reversing the antiferroaxial order deterministically switches the spin splitting and time-reversal-odd responses such as AHC.

Our results establish antiferroaxiality as a key ferroic degree of freedom enabling the emergence and active control of altermagnetism. Recent progress in ferroaxial materials has demonstrated deterministic switching of rotational order using circularly polarized terahertz pulses via nonlinear phonon excitation~\cite{heOpticalControlFerroaxial2024,zengPhotoinducedNonvolatileRewritable2025}.
Motivated by this progress, and considering that antiferroaxial order arises from staggered rotational lattice modes, the analogous dynamical protocols could be extended to manipulate antiferroaxiality.
Importantly, in the present framework, reversing the antiferroaxial order not only flips the altermagnetic spin splitting and associated transport responses, but also constitutes a deterministic reversal of the underlying structural order itself.
Furthermore, we show that antiferroaxiality can be efficiently tuned and switched on or off by uniaxial strain, and identify KMnF$_3$ as a concrete material realization (see SM for details~\cite{SM}).
This work would establish a new paradigm for altermagnetic spintronics, enabling structural-magnetic co-engineering for the active control of spintronic devices even in centrosymmetric materials.

\begin{acknowledgments}
\textit{Acknowledgments.}--- The work is supported by the Science Fund for Creative Research Groups of NSFC (Grant No. 12321004), the NSF of China (Grant No. 12374055), and the National Key R\&D Program of China (Grant No. 2020YFA0308800).
\end{acknowledgments}

\bibliography{ref}
\appendix

\section*{End matter}
In the zero SOC limit, a collinear altermagnet is characterized by a N\'{e}el vector $\mathbf N$ that transforms as
$\Gamma_{\rm AM}\otimes\Gamma_A^s$.
Here $\Gamma_{\rm AM}$ is a nontrivial 1D IR of the crystallographic point group and, when inversion is present, it must be inversion-even, while $\Gamma_A^s$ denotes the axial vector IR in spin-rotation space.
This transformation property implies that neither $\mathcal T\boldsymbol\tau$ nor $\mathcal I\mathcal T$ enforces spin degeneracy in the zero SOC limit, allowing a momentum-dependent spin splitting.
Given the altermagnetic spin point group, we determine whether the ordered phase is altermagnetic and identify whether its spin splitting belongs to the $d$-wave, $g$-wave, or $i$-wave type.
In the present work, once the antiferroaxial and N\'{e}el orders select a specific $\Gamma_{\rm AM}$ through the symmetry-allowed coupling, Table~\ref{table:TB1} provides a direct lookup of the corresponding spin splitting type.
All IRs with allowed nontrivial spin splitting are listed in Table~\ref{table:TB1}. A more detailed classification, including the associated lowest-order nodal functions, is provided in Table~S3 in SM~\cite{SM}

It is worth noting that, in the parent group $\mathcal{G}_0$, the bilinear product transforms as $\mathbf{G}\mathbf{N}\sim \Gamma_G(\mathbf{q})\otimes\Gamma_A \otimes\Gamma_N(\mathbf{q}')\otimes\Gamma_A^s$.
After $\mathbf{G}$ condenses, the symmetry is lowered to the isotropy subgroup $\mathcal{G}$, and the order parameter $\mathbf{G}$ transforms as the identity representation, that is $(\Gamma_G(\mathbf{q})\otimes \Gamma_A)\downarrow \mathcal{G}=\Gamma_1$, and the N\'{e}el vector transforms according to the altermagnetic IR $\Gamma_{\mathrm{AM}}\otimes\Gamma_A^s$, namely $(\Gamma_N(\mathbf{q}')\otimes\Gamma_A^s)\downarrow \mathcal{G}=\Gamma_{\mathrm{AM}}\otimes\Gamma_A^s$. Accordingly, Eq.~(\ref{Eq:AM_quest}) can be stated equivalently as follows: a system realizes antiferroaxial altermagnetism if and only if, upon condensation of the antiferroaxial order, the irreducible representation of the N\'{e}el vector in the resulting isotropy subgroup belongs to one of the $\Gamma_{\rm AM}$ listed in Table~\ref{table:TB1}.

\begin{table}[b]
\caption{\label{table:TB1}
Symmetry classification of antiferroaxial altermagnetism by collinear spin point groups (SPG).
The table lists crystallographic point groups (PG) and their symmetry-allowed collinear spin point groups.
$\Gamma_\text{AM}$ denotes the irreducible representation of the N\'{e}el vector.
The spin splitting type is categorized into $d$-wave ($d$), planar/bulk $g$-wave ($g_{p/b}$), and planar/bulk $i$-wave ($i_{p/b}$).
The subscripts $\alpha \in \{1,2,3\}$ and $\beta \in \{1,2\}$ denote the distinct irreducible representations corresponding to the same spin point group.
Mulliken symbols are used for the irreducible representations of point groups, while spin point groups follow Litvin notation~\cite{litvinSpinPointGroups1977}.
Groups with no allowed non-trivial spin splitting are omitted.}
\footnotesize
\renewcommand{\arraystretch}{1.25}
\setlength{\tabcolsep}{4pt}
\begin{ruledtabular}
\begin{tabular}{cl cl}
PG & SPG $(\Gamma_\text{AM}, \text{Type})$ & PG & SPG $(\Gamma_\text{AM}, \text{Type})$ \\
\colrule
$2$   & $^{\bar{1}}2$ ($B$, $d$)                & $32$ & $3^{\bar{1}}2$ ($A_2$, $g_b$) \\
$m$   & $^{\bar{1}}m$ ($A^{\prime\prime}$, $d$) & $3m$ & $3^{\bar{1}}m$ ($A_2$, $g_b$) \\
$2/m$ & $^{\bar{1}}2/^{\bar{1}}m$ ($B_g$, $d$)  & $\bar{3}m$ & $\bar{3}^{\bar{1}}m$ ($A_{2g}$, $g_b$) \\
$222$ & $2^{\bar{1}}2^{\bar{1}}2$ ($B_{\alpha}$, $d$) & $6$  & $^{\bar{1}}6$ ($B$, $g_b$) \\
$mm2$ & $^{\bar{1}}m^{\bar{1}}m2$ ($A_2$, $d$)   & $\bar{6}$ & $^{\bar{1}}\bar{6}$ ($A^{\prime\prime}$, $g_b$) \\
      & $m^{\bar{1}}m^{\bar{1}}2$ ($B_{\beta}$, $d$) & $6/m$ & $^{\bar{1}}6/^{\bar{1}}m$ ($B_g$, $g_b$) \\
$mmm$ & $m^{\bar{1}}m^{\bar{1}}m$ ($B_{\alpha g}$, $d$) & $622$ & $^{\bar{1}}62^{\bar{1}}2$ ($B_{\beta}$, $g_b$) \\
$4$   & $^{\bar{1}}4$ ($B$, $d$)                &       & $6^{\bar{1}}2^{\bar{1}}2$ ($A_2$, $i_p$) \\
$\bar{4}$ & $^{\bar{1}}\bar{4}$ ($B$, $d$)      & $6mm$ & $^{\bar{1}}6m^{\bar{1}}m$ ($B_{\beta}$, $g_b$) \\
$4/m$ & $^{\bar{1}}4/m$ ($B_g$, $d$)            &       & $6^{\bar{1}}m^{\bar{1}}m$ ($A_2$, $i_p$) \\
$422$ & $^{\bar{1}}42^{\bar{1}}2$ ($B_{\beta}$, $d$) & $\bar{6}m2$ & $^{\bar{1}}\bar{6}m^{\bar{1}}2$ ($A_2^{\prime\prime}$, $g_b$) \\
      & $4^{\bar{1}}2^{\bar{1}}2$ ($A_2$, $g_p$) &             & $^{\bar{1}}\bar{6}^{\bar{1}}m2$ ($A_1^{\prime\prime}$, $g_b$) \\
$4mm$ & $^{\bar{1}}4m^{\bar{1}}m$ ($B_{\beta}$, $d$) &             & $\bar{6}^{\bar{1}}m^{\bar{1}}2$ ($A_2^\prime$, $i_p$) \\
      & $4^{\bar{1}}m^{\bar{1}}m$ ($A_2$, $g_p$) & $6/mmm$     & $^{\bar{1}}6/^{\bar{1}}mm^{\bar{1}}m$ ($B_{\beta g}$, $g_b$) \\
$\bar{4}2m$ & $^{\bar{1}}\bar{4}^{\bar{1}}2m$ ($B_2$, $d$) &         & $6/m^{\bar{1}}m^{\bar{1}}m$ ($A_{2g}$, $i_p$) \\
            & $^{\bar{1}}\bar{4}2^{\bar{1}}m$ ($B_1$, $d$) & $\bar{4}3m$ & $^{\bar{1}}\bar{4}3^{\bar{1}}m$ ($A_2$, $i_b$) \\
            & $\bar{4}^{\bar{1}}2^{\bar{1}}m$ ($A_2$, $g_p$) & $432$       & $^{\bar{1}}43^{\bar{1}}2$ ($A_2$, $i_b$) \\
$4/mmm$     & $^{\bar{1}}4/m^{\bar{1}}mm$ ($B_{\beta g}$, $d$) & $m\bar{3}m$ & $^{\bar{1}}m\bar{3}^{\bar{1}}m$ ($A_{2g}$, $i_b$) \\
            & $4/m^{\bar{1}}m^{\bar{1}}m$ ($A_{2g}$, $g_p$) &             & \\
\end{tabular}
\end{ruledtabular}
\end{table}

\end{document}


\title{Supplementary material for ``Antiferroaxial altermagnetism"}
\author{Yichen Liu$^{1}$}
\author{Cheng-Cheng Liu$^{1}$}
\email{ccliu@bit.edu.cn}
\affiliation{$^{1}$Centre for Quantum Physics, Key Laboratory of Advanced Optoelectronic Quantum Architecture and Measurement (MOE), School of Physics, Beijing Institute of Technology, Beijing 100081, China}

\maketitle

This Supplementary Material provides technical details and supporting data for the main text.
In Sec.~\ref{sec:1}, we employ symmetry analysis to demonstrate how antiferroaxial order gives rise to $d$-, $g$-, and $i$-wave altermagnetism.
Section~\ref{sec:AM_Neel_SPG_mapping} classifies antiferroaxial altermagnetism.
Section~\ref{sec:isotropy} identifies the isotropy subgroup and derives the subduced representation of the Néel vector.
In Sec.~\ref{sec:2}, we detail the ligand-rotation tight-binding models and present their rigorous derivation, symmetry analysis and band structures.
Section~\ref{sec:3} presents a Landau free energy analysis for space group $Pm\bar{3}m$, illustrated by FeF$_3$ and KMnF$_3$.
Finally, Sec.~\ref{sec:material} provides computational details for the density functional theory calculations and candidate antiferroaxial altermagnets, such as MnS$_2$, MnC$_2$, FeBr$_3$, KMnF$_3$, Ca$_2$MnO$_4$, FeF$_3$, CoF$_3$, LaCrO$_3$, and La$_2$NiO$_4$.

\section{Possible magnetic multipoles in antiferroaxial altermagnetism}\label{sec:1}

In this section, we classify the magnetic multipoles induced by coexisting antiferroaxial and N\'{e}el orders. This classification determines the symmetry allowed antiferroaxial altermagnetic spin splitting types in each crystallographic point group.
We work in the limit of zero spin-orbit coupling (SOC), where the spin and real-space sectors are separable.

In the parent space group $\mathcal G_0$, an antiferroaxial order parameter $\mathbf G$ transforms as
$\mathbf G \sim \Gamma_G(\mathbf q)\otimes \Gamma_A$,
where $\Gamma_G(\mathbf q)$ is an irreducible representation (IR) of the little group at propagation vector $\mathbf q$ that encodes the staggered spatial modulation, and $\Gamma_A$ denotes the axial-vector representation of the crystallographic point group.
Similarly, the N\'{e}el vector transforms as
$\mathbf N \sim \Gamma_N(\mathbf q')\otimes \Gamma_A^{s}$,
where $\Gamma_N(\mathbf q')$ is an IR of the little group at $\mathbf q'$ and $\Gamma_A^{s}$ is the axial-vector representation in spin-rotation space.

The nonrelativistic spin splitting in altermagnets can be parameterized by magnetic multipoles, defined as rank-$\ell$ symmetric-traceless moments of the magnetization density $\mathbf m(\mathbf r)$,
\begin{equation}
\mathbf O^{(\ell)} \equiv \int d^3r \,\big[r_{\mu_1}r_{\mu_2}\cdots r_{\mu_\ell}\big]\mathbf m(\mathbf r),
\label{eq:Omultipole_SM}
\end{equation}
where $r_\mu$ are Cartesian components of $\mathbf r$ and $[\cdots]$ denotes the symmetric-traceless projection over $\mu_1\cdots\mu_\ell$.
By construction, $\mathbf O^{(\ell)}$ transforms as $\Gamma_\ell\otimes \Gamma_A^{s}$, where $\Gamma_\ell$ is the spatial rank-$\ell$ representation (dimension $2\ell+1$) that subduces into IRs of the crystallographic point group.
The ranks $\ell=2,4,6$ correspond to the $d$-, $g$-, and $i$-wave altermagnetic symmetries, respectively.

We consider a symmetry-allowed trilinear invariant coupling among $\mathbf G$, $\mathbf N$, and $\mathbf O^{(\ell)}$. In the parent phase, the identity representation must preserve the translational symmetry of the parent group, which enforces momentum conservation, $\mathbf q + \mathbf q' = \mathbf K$ (where $\mathbf K$ is a reciprocal lattice vector); i.e., $\mathbf q' = -\mathbf q$ up to a reciprocal lattice vector.
Since $\Gamma_1^{s}\subset \Gamma_A^{s}\otimes\Gamma_A^{s}$, the invariance condition reduces to the real-space sector
\begin{equation}
\Gamma_1 \subset \Gamma_G(\mathbf q)\otimes \Gamma_A \otimes \Gamma_N(\mathbf q') \otimes \Gamma_\ell .
\label{eq:tri_invariant_SM}
\end{equation}
Hereafter, we suppress the spin-rotation factor $\Gamma_A^s$ for notational brevity. It is convenient to introduce the combined modulation representation
$\Gamma_0 \equiv \Gamma_G(\mathbf q)\otimes \Gamma_N(\mathbf q')$,
so that the bilinear coupling transforms as $\mathbf N \mathbf G\sim \Gamma_0\otimes \Gamma_A$.
Equation~(\ref{eq:tri_invariant_SM}) then reduces to a purely spatial selection rule: a trilinear invariant exists if and only if $\Gamma_0\otimes\Gamma_A$ and the multipole sector $\Gamma_\ell$ share at least one common point-group IR, i.e.,
\begin{equation}
\big(\Gamma_0\otimes\Gamma_A\big)\cap \Gamma_\ell \neq \emptyset .
\end{equation}
Therefore, once $\Gamma_G(\mathbf q)$ and $\Gamma_N(\mathbf q')$ are specified in the parent group, this criterion fixes the lowest allowed multipole rank $\ell_{\min}$ (and hence the $d$/$g$/$i$-wave class), while the detailed basis form (e.g., $d_{xy}$ versus $d_{x^2-y^2}$) is only resolved after the order parameter condenses and the system reduces to the isotropy subgroup.

In practice, for each crystallographic point group, we:
(i) specify the real-space axial-vector representation $\Gamma_A$;
(ii) enumerate admissible $\Gamma_0$ compatible with a collinear N\'{e}el order in that point group (for centrosymmetric groups, $\Gamma_0$ must be inversion-even for the even-$\ell$ multipoles considered here);
(iii) tabulate $\Gamma_0\otimes\Gamma_A$; and
(iv) determine the lowest $\ell=2,4,6$ whose multipole decomposition contains at least one IR that also appears in the decomposition of $\Gamma_0\otimes\Gamma_A$.
The decompositions of $\Gamma_\ell$ into point-group IRs are listed in Table~\ref{table:TB1}, and the resulting allowed spin-splitting types are summarized in Table~\ref{table:TB2}.

This analysis shows that antiferroaxiality generically enables $d$-, $g$-, and $i$-wave altermagnetic spin splitting depending on the point-group symmetry and the specific $\Gamma_0$ channel selected by the antiferroaxial--N\'{e}el coexistence.

\begin{table*}[t]
\caption{\label{table:TB1}
Decomposition of magnetic multipoles under crystallographic point groups.
The table lists the irreducible representations (IRs) contained in the spatial sector $\Gamma_\ell$ of the magnetic multipoles for rank $\ell=2$ ($d$-wave altermagnetism), $\ell=4$ ($g$-wave altermagnetism), and $\ell=6$ ($i$-wave altermagnetism).
The notation follows the Mulliken symbols.}
\begin{ruledtabular}
\begin{tabular}{cccc}
Point Group & $d$-wave ($\ell=2$) & $g$-wave ($\ell=4$) & $i$-wave ($\ell=6$)\\
\colrule
$\bar{1}$ & $5 A_u$ & $9A_u$ & $13 A_u$\\
2 & $3A\oplus 2B$ & $5 A\oplus 4B$& $7A\oplus 6B$\\
m & $3A'\oplus 2A''$ & $5 A'\oplus 4A''$& $7A'\oplus 6A''$\\
2/m & $3A_g\oplus 2B_g$ & $5 A_g\oplus 4 B_g$ & $7A_g\oplus 6B_g$\\
222 & $2A\oplus B_1\oplus B_2\oplus B_3$ & $3A\oplus 2B_1\oplus 2B_2\oplus 2B_3$ &$4A\oplus 3B_1\oplus 3B_2\oplus 3B_3$\\
mm2 & $2A_1\oplus A_2\oplus B_1\oplus B_2$ & $3A_1\oplus 2A_2\oplus 2B_1\oplus 2B_2$ & $4A_1\oplus 3A_2\oplus 3B_1\oplus 3B_2$ \\
mmm & $2A_g\oplus B_{1g}\oplus B_{2g}\oplus B_{3g}$ & $3A_g\oplus 2 B_{1g}\oplus 2B_{2g}\oplus 2B_{3g}$ & $4A_g\oplus 3B_{1g}\oplus 3B_{2g}\oplus 3B_{3g}$\\
4, $\bar{4}$ & $A\oplus 2B\oplus E$ & $3A\oplus 2B\oplus 2E$ & $3A\oplus 4B\oplus 3E$\\
4/m & $A_g\oplus 2B_g\oplus E_g$ & $3A_g\oplus 2B_g\oplus 2E_g$ & $3A_g\oplus 4B_g\oplus 3E_g$\\
422, 4mm, $\bar{4}$2m & $A_1\oplus B_1\oplus B_2\oplus E$ & $2A_1\oplus A_2\oplus B_1\oplus B_2\oplus 2E$ & $2A_1\oplus A_2\oplus 2B_1\oplus 2B_2\oplus 3E$ \\
4/mmm & $A_{1g}\oplus B_{1g}\oplus B_{2g}\oplus E_g$ & $2A_{1g}\oplus A_{2g}\oplus B_{1g}\oplus B_{2g}\oplus 2E_g$ & $2A_{1g}\oplus A_{2g}\oplus 2B_{1g}\oplus 2B_{2g}\oplus 3E_g$\\
3 & 2$E$ & 3$E$ & 4 $E$\\
$\bar{3}$ & $A_{g}\oplus 2E_g$ & $3A_{g}\oplus 3E_g$ & $5A_g\oplus 4E_g$\\
32, 3m & $A_1\oplus 2E$ & $2A_1\oplus A_2\oplus 3E$ & $3A_1\oplus 2A_2\oplus 4E$\\
$\bar{3}$m & $A_{1g}\oplus 2E_g$ & $2A_{1g}\oplus A_{2g}\oplus 3E_g$ & $3A_{1g}\oplus 2A_{2g}\oplus 4 E_g$\\
6 & $A\oplus E_1\oplus E_2$ & $A\oplus 2B\oplus E_1\oplus 2E_2$ &  $3A\oplus 2B\oplus 2E_1\oplus 2E_2$ \\
$\bar{6}$ & $A'\oplus E'\oplus E''$ & $2A'\oplus A''\oplus 2E'\oplus E''$ & $3A'\oplus 2A''\oplus 2E'\oplus 2E''$\\
6/m & $A_g\oplus E_{1g}\oplus E_{2g}$ & $A_g\oplus 2B_{g}\oplus E_{1g}\oplus 2E_{2g}$ & $3A_g\oplus 2B_g\oplus 2E_{1g}\oplus 2E_{2g}$\\
622, 6mm & $A_1\oplus E_1\oplus E_2$ & $A_1\oplus B_1\oplus B_2\oplus E_1\oplus 2E_2$ & $2A_1\oplus A_2\oplus B_1\oplus B_2\oplus E_1\oplus 2E_2$\\
$\bar{6}$m2 & $A_1'\oplus E'\oplus E''$ & $A_1'\oplus A_1''\oplus A_2''\oplus E''\oplus 2E'$ & $2A_1'\oplus A_2'\oplus A_1''\oplus A_2''\oplus 2E'\oplus 2E''$\\
6/mmm & $A_{1g}\oplus E_{1g}\oplus E_{2g}$ & $A_{1g}\oplus B_{1g}\oplus B_{2g}\oplus E_{1g}\oplus 2E_{2g}$ & $2A_{1g}\oplus A_{2g}\oplus B_{1g}\oplus B_{2g}\oplus 2E_{1g}\oplus 2E_{2g}$\\
23 & $E\oplus T$ & $A\oplus E\oplus 2T$ & $2A\oplus E\oplus 2T$\\
m$\bar{3}$ & $E_g\oplus T_g$ & $A_g\oplus E_g\oplus 2T_g$ & $2A_g\oplus E_g\oplus 3T_g$\\
432, $\bar{4}$3m & $E\oplus T_2$ & $A_1\oplus E\oplus T_1\oplus T_2$ & $A_1\oplus A_2\oplus E\oplus T_1\oplus 2T_2$\\
m$\bar{3}$m & $E_g\oplus T_{2g}$ & $A_{1g}\oplus E_g\oplus T_{1g}\oplus T_{2g}$ & $A_{1g}\oplus A_{2g}\oplus E_g\oplus T_{1g}\oplus 2T_{2g}$\\
\end{tabular}
\end{ruledtabular}
\end{table*}

\begin{table*}[h]
\caption{\label{table:TB2}
Symmetry criteria for antiferroaxial altermagnetism.
The table lists the axial vector representation $\Gamma_A$ and the product representation $\Gamma_0\otimes\Gamma_A$ for each point group.
$\Gamma_0$ is an inversion-even IR representing the combined spatial parity of the antiferroaxial and N\'{e}el orders.
The last column indicates the allowed altermagnetic spin splitting types ($d$-, $g$-, or $i$-wave) derived from the overlap with Table~\ref{table:TB1}.}
\footnotesize
\newcommand{\splitcell}[2]{\begin{tabular}{@{}c@{}}#1\\#2\end{tabular}}

\begin{ruledtabular}
\begin{tabular}{cccc}
Point Group & $\Gamma_A$  & Possible $\Gamma_0\otimes\Gamma_A$ & Allowed Spin Splitting\\
\colrule
$\bar{1}$ & $3A_g$ & $3A_g$ & $\emptyset$\\
2 & $A\oplus 2B$ & $A\oplus 2B$, $B\oplus 2A$ & $d$\\
m & $A'\oplus 2A''$ & $A'\oplus 2A''$, $A''\oplus 2A'$ & $d$\\
2/m & $A_g\oplus 2B_g$ & $A_g\oplus 2B_g$, $B_g\oplus 2A_g$ & $d$\\
222 & $B_1\oplus B_2\oplus B_3$ & $B_1\oplus B_2\oplus B_3$, $A\oplus B_2\oplus B_3$, $A\oplus B_1\oplus B_3$, $A\oplus B_1\oplus B_2$ & $d$\\
mm2 & $A_2\oplus B_1\oplus B_2$ & $A_2\oplus B_1\oplus B_2$, $A_1\oplus B_1\oplus B_2$, $A_1\oplus A_2\oplus B_2$, $A_1\oplus A_2\oplus B_1$ & $d$\\
mmm & $B_{1g}\oplus B_{2g}\oplus B_{3g}$ & $B_{1g}\oplus B_{2g}\oplus B_{3g}$, $A_g\oplus B_{2g}\oplus B_{3g}$, $A_g\oplus B_{1g}\oplus B_{3g}$ & $d$\\
4 & $A\oplus E$ & $A\oplus E$, $B\oplus E$, $2A\oplus 2B\oplus E$ & $d$\\
$\bar{4}$ & $A\oplus E$ & $A\oplus E$, $B\oplus E$, $2A\oplus 2B\oplus E$ & $d$\\
$4/m$ & $A_g\oplus E_g$ & $A_g\oplus E_g$, $B_g\oplus E_g$, $2A_g\oplus 2B_g\oplus E_g$ & $d$\\
422 & $A_2\oplus E$ & \splitcell{$A_2\oplus E, A_1\oplus E, B_2\oplus E, B_1\oplus E$,}{$A_1\oplus A_2\oplus B_1\oplus B_2\oplus E$} & $d,g$\\
4mm & $A_2\oplus E$ & \splitcell{$A_2\oplus E, A_1\oplus E, B_2\oplus E, B_1\oplus E$,}{$A_1\oplus A_2\oplus B_1\oplus B_2\oplus E$} & $d,g$\\
$\bar{4}2m$ & $A_2\oplus E$ & \splitcell{$A_2\oplus E, A_1\oplus E, B_1\oplus E, B_2\oplus E$,}{$A_1\oplus A_2\oplus B_1\oplus B_2\oplus E$} & $d,g$\\
4/mmm & $A_{2g}\oplus E_g$ & \splitcell{$A_{2g}\oplus E_g, A_{1g}\oplus E_g, B_{1g}\oplus E_g, B_{2g}\oplus E_g$,}{$A_{1g}\oplus A_{2g}\oplus B_{1g}\oplus B_{2g}\oplus E_g$} & $d,g$\\
3 & $A\oplus E$ & $A\oplus E$, $2A\oplus 2E$ & $d$\\
$\bar{3}$ & $A_g\oplus E_g$ & $A_g\oplus E_g$, $2A_g\oplus 2E_g$ & $d$\\
32 & $A_2\oplus E$ & $A_2\oplus E$, $A_1\oplus E$, $A_1\oplus A_2\oplus 2E$ & $d,g$\\
3m & $A_2\oplus E$ & $A_2\oplus E$, $A_1\oplus E$, $A_1\oplus A_2\oplus 2E$ & $d,g$\\
$\bar{3}m$ & $A_{2g}\oplus E_g$ & $A_{2g}\oplus E_g$, $A_{1g}\oplus E_g$, $A_{1g}\oplus A_{2g}\oplus 2E_g$ & $d,g$\\
6 & $A\oplus E_1$ & $A\oplus E_1$, $B\oplus E_2$, $2A\oplus E_1\oplus E_2$, $2B\oplus E_1\oplus E_2$ & $d,g$\\
$\bar{6}$ & $A'\oplus E'$ & $A'\oplus E'$, $A''\oplus E''$, $2A'\oplus2 E'$, $2A''\oplus2 E''$ & $d,g$\\
$6/m$ & $A_g\oplus E_{1g}$ & $A_g\oplus E_{1g}$, $B_g\oplus E_{2g}$, $2A_g\oplus E_{1g}\oplus E_{2g}$, $2B_g\oplus E_{1g}\oplus E_{2g}$  & $d,g$\\
622 & $A_2\oplus E_1$ & \splitcell{$A_2\oplus E_1, A_1\oplus E_1, B_2\oplus E_2, B_1\oplus E_2$,}{$A_1\oplus A_2\oplus E_1\oplus E_2, B_1\oplus B_2\oplus E_1\oplus E_2$} & $d,g,i$\\
6mm & $A_2\oplus E_1$ & \splitcell{$A_2\oplus E_1, A_1\oplus E_1, B_2\oplus E_2, B_1\oplus E_2$,}{$A_1\oplus A_2\oplus E_1\oplus E_2, B_1\oplus B_2\oplus E_1\oplus E_2$}  & $d,g,i$\\
$\bar{6}m2$ & $A_2^\prime\oplus E^{\prime\prime}$ & \splitcell{$A_2^\prime\oplus E^{\prime\prime}, A_1^\prime\oplus E^{\prime\prime}, A_2^{\prime\prime}\oplus E^\prime, A_1^{\prime\prime}\oplus E^\prime$,}{$A_1^{\prime\prime}\oplus A_2^{\prime\prime}\oplus E^\prime\oplus E^{\prime\prime}, A_1^{\prime}\oplus A_2^{\prime}\oplus E^\prime\oplus E^{\prime\prime}$} & $d,g,i$\\
$6$/mmm & $A_{2g}\oplus E_{1g}$ & \splitcell{$A_{2g}\oplus E_{1g}, A_{1g}\oplus E_{1g}, B_{2g}\oplus E_{2g}, B_{1g}\oplus E_{2g}$,}{$A_{1g}\oplus A_{2g}\oplus E_{1g}\oplus E_{2g}, B_{1g}\oplus B_{2g}\oplus E_{1g}\oplus E_{2g}$} & $d,g,i$\\
23 & $T$ & $T$, $2 T$, $A\oplus E\oplus 2 T$ & $d$\\
m$\bar{3}$ & $T_g$ & $T_g$ , $2 T_g$, $A_g\oplus E_g\oplus 2 T_g$& $d$\\
432 & $T_1$ & \splitcell{$T_1, T_2, T_1\oplus T_2$,}{$A_2\oplus E\oplus T_1\oplus T_2, A_1\oplus E\oplus T_1\oplus T_2$} & $d,g,i$\\
$\bar{4}3m$ & $T_1$ & \splitcell{$T_1, T_2, T_1\oplus T_2$,}{$A_2\oplus E\oplus T_1\oplus T_2, A_1\oplus E\oplus T_1\oplus T_2$} & $d,g,i$\\
m$\bar{3}$m & $T_{1g}$ & \splitcell{$T_{1g}, T_{2g}, T_{1g}\oplus T_{2g}$,}{$A_{1g}\oplus E_G\oplus T_{1g}\oplus T_{2g}, A_{2g}\oplus E_G\oplus T_{1g}\oplus T_{2g}$} & $d,g,i$\\
\end{tabular}
\end{ruledtabular}
\end{table*}

\section{Classification of antiferroaxial altermagnetism in terms of spin point group}\label{sec:AM_Neel_SPG_mapping}

In the zero SOC limit, the altermagnetism is captured in Landau theory by requiring that the N\'{e}el vector transforms as $\mathbf N \sim \Gamma_{\mathrm{AM}}\otimes \Gamma_A^s$, where $\Gamma_{\mathrm{AM}}$ is a nontrivial, inversion-even, one-dimensional irreducible representation (IR) of the crystallographic point group, and $\Gamma_A^s$ denotes the axial-vector IR in spin-rotation space.
The nonrelativistic spin splitting is conveniently characterized by a magnetic multipole, $\mathbf O^{(\ell)} \sim \Gamma_\ell \otimes \Gamma_A^s$, which appears as a secondary order parameter linearly coupled to altermagnetic N\'{e}el vector $\mathbf N$.
Accordingly, the type of altermagnetism is determined by the lowest-rank multipole sector (smallest $\ell$) that is symmetry-allowed to couple to the N\'{e}el order, with $\ell=2,4,6$ corresponding to the $d$-, $g$-, and $i$-wave altermagnetism, respectively. Table~\ref{table:altermagnet_valid_only} provides a comprehensive classification of these symmetry-allowed antiferroaxial altermagnetic phases, listing the corresponding spin point groups, irreducible representations, and lowest-order nodal functions for all relevant crystallographic point groups.

\begin{table*}[htbp]
\centering
\caption{\label{table:altermagnet_valid_only}Classification of antiferroaxial altermagnetism. This table lists only the symmetry-allowed altermagnetic orders, excluding ferromagnetic (FM) and trivial non-splitting cases. It provides the point group (PG), the specific collinear spin point group (SPG) notation reflecting the preserved spatial axes, the corresponding irreducible representation ($\Gamma_\text{AM}$), and the spin splitting type with the lowest-order nodal function.}
\footnotesize
\begin{ruledtabular}
\begin{tabular}{cll @{\hspace{0.8cm}} cll}
\textbf{PG} & \textbf{SPG ($\Gamma_\text{AM}$)} & \textbf{Splitting [Function]} &
\textbf{PG} & \textbf{SPG ($\Gamma_\text{AM}$)} & \textbf{Splitting [Function]} \\
\colrule
$2$       & $^{\bar{1}}2$ ($B$) & $d$-wave $[xy]$ &
$4/mmm$   & $4/m^{\bar{1}}m^{\bar{1}}m$ ($A_{2g}$) & $g$-wave $[xy(x^2-y^2)]$ \\
$m$       & $^{\bar{1}}m$ ($A''$) & $d$-wave $[xy]$ &
          & $^{\bar{1}}4/mm^{\bar{1}}m$ ($B_{1g}$) & $d$-wave $[x^2-y^2]$ \\
$2/m$     & $^{\bar{1}}2/^{\bar{1}}m$ ($B_g$) & $d$-wave $[xy]$ &
          & $^{\bar{1}}4/m^{\bar{1}}mm$ ($B_{2g}$) & $d$-wave $[xy]$ \\
$222$     & $^{\bar{1}}2^{\bar{1}}22$ ($B_1$) & $d$-wave $[xy]$ &
$32$      & $3^{\bar{1}}2$ ($A_2$) & $g$-wave $[z(x^3-3xy^2)]$ \\
          & $^{\bar{1}}22^{\bar{1}}2$ ($B_2$) & $d$-wave $[xz]$ &
$3m$      & $3^{\bar{1}}m$ ($A_2$) & $g$-wave $[z(x^3-3xy^2)]$ \\
          & $2^{\bar{1}}2^{\bar{1}}2$ ($B_3$) & $d$-wave $[yz]$ &
$\bar{3}m$& $\bar{3}^{\bar{1}}m$ ($A_{2g}$) & $g$-wave $[z(x^3-3xy^2)]$ \\
$mm2$     & $^{\bar{1}}m^{\bar{1}}m2$ ($A_2$) & $d$-wave $[xy]$ &
$6$       & $^{\bar{1}}6$ ($B$) & $g$-wave $[z(x^3-3xy^2)]$ \\
          & $m^{\bar{1}}m^{\bar{1}}2$ ($B_1$) & $d$-wave $[xz]$ &
$\bar{6}$ & $^{\bar{1}}\bar{6}$ ($A''$) & $g$-wave $[z(x^3-3xy^2)]$ \\
          & $^{\bar{1}}mm^{\bar{1}}2$ ($B_2$) & $d$-wave $[yz]$ &
$6/m$     & $^{\bar{1}}6/^{\bar{1}}m$ ($B_g$) & $g$-wave $[z(x^3-3xy^2)]$ \\
$mmm$     & $^{\bar{1}}m^{\bar{1}}mm$ ($B_{1g}$) & $d$-wave $[xy]$ &
$622$     & $6^{\bar{1}}2^{\bar{1}}2$ ($A_2$) & $i$-wave $[xy(3x^2-y^2)(x^2-3y^2)]$ \\
          & $^{\bar{1}}mm^{\bar{1}}m$ ($B_{2g}$) & $d$-wave $[xz]$ &
          & $^{\bar{1}}62^{\bar{1}}2$ ($B_1$) & $g$-wave $[z(3x^2y-y^3)]$ \\
          & $m^{\bar{1}}m^{\bar{1}}m$ ($B_{3g}$) & $d$-wave $[yz]$ &
          & $^{\bar{1}}6^{\bar{1}}22$ ($B_2$) & $g$-wave $[z(x^3-3xy^2)]$ \\
$4$       & $^{\bar{1}}4$ ($B$) & $d$-wave $[x^2-y^2, xy]$ &
$6mm$     & $6^{\bar{1}}m^{\bar{1}}m$ ($A_2$) & $i$-wave $[xy(3x^2-y^2)(x^2-3y^2)]$ \\
$\bar{4}$ & $^{\bar{1}}\bar{4}$ ($B$) & $d$-wave $[x^2-y^2, xy]$ &
          & $^{\bar{1}}6m^{\bar{1}}m$ ($B_1$) & $g$-wave $[z(3x^2y-y^3)]$ \\
$4/m$     & $^{\bar{1}}4/m$ ($B_g$) & $d$-wave $[x^2-y^2, xy]$ &
          & $^{\bar{1}}6^{\bar{1}}mm$ ($B_2$) & $g$-wave $[z(x^3-3xy^2)]$ \\
$422$     & $4^{\bar{1}}2^{\bar{1}}2$ ($A_2$) & $g$-wave $[xy(x^2-y^2)]$ &
$\bar{6}m2$& $\bar{6}^{\bar{1}}m^{\bar{1}}2$ ($A_2'$) & $i$-wave $[xy(3x^2-y^2)(x^2-3y^2)]$ \\
          & $^{\bar{1}}42^{\bar{1}}2$ ($B_1$) & $d$-wave $[x^2-y^2]$ &
          & $^{\bar{1}}\bar{6}^{\bar{1}}m2$ ($A_1''$) & $g$-wave $[z(3x^2y-y^3)]$ \\
          & $^{\bar{1}}4^{\bar{1}}22$ ($B_2$) & $d$-wave $[xy]$ &
          & $^{\bar{1}}\bar{6}m^{\bar{1}}2$ ($A_2''$) & $g$-wave $[z(x^3-3xy^2)]$ \\
$4mm$     & $4^{\bar{1}}m^{\bar{1}}m$ ($A_2$) & $g$-wave $[xy(x^2-y^2)]$ &
$6/mmm$   & $6/m^{\bar{1}}m^{\bar{1}}m$ ($A_{2g}$) & $i$-wave $[xy(3x^2-y^2)(x^2-3y^2)]$ \\
          & $^{\bar{1}}4m^{\bar{1}}m$ ($B_1$) & $d$-wave $[x^2-y^2]$ &
          & $^{\bar{1}}6/mm^{\bar{1}}m$ ($B_{1g}$) & $g$-wave $[z(3x^2y-y^3)]$ \\
          & $^{\bar{1}}4^{\bar{1}}mm$ ($B_2$) & $d$-wave $[xy]$ &
          & $^{\bar{1}}6/m^{\bar{1}}mm$ ($B_{2g}$) & $g$-wave $[z(x^3-3xy^2)]$ \\
$\bar{4}2m$& $\bar{4}^{\bar{1}}2^{\bar{1}}m$ ($A_2$) & $g$-wave $[xy(x^2-y^2)]$ &
$432$     & $^{\bar{1}}43^{\bar{1}}2$ ($A_2$) & $i$-wave $[(x^2-y^2)(y^2-z^2)(z^2-x^2)]$ \\
          & $^{\bar{1}}\bar{4}2^{\bar{1}}m$ ($B_1$) & $d$-wave $[x^2-y^2]$ &
$\bar{4}3m$& $^{\bar{1}}\bar{4}3^{\bar{1}}m$ ($A_2$) & $i$-wave $[(x^2-y^2)(y^2-z^2)(z^2-x^2)]$ \\
          & $^{\bar{1}}\bar{4}^{\bar{1}}2m$ ($B_2$) & $d$-wave $[xy]$ &
$m\bar{3}m$& $^{\bar{1}}4/m\bar{3}^{\bar{1}}2/^{\bar{1}}m$ ($A_{2g}$) & $i$-wave $[(x^2-y^2)(y^2-z^2)(z^2-x^2)]$ \\
\end{tabular}
\end{ruledtabular}
\end{table*}

\section{Isotropy subgroup and irrep subduction}\label{sec:isotropy}
The isotropy subgroup $\mathcal G$~\cite{hatch1989isotropy} associated with the condensation of the antiferroaxial order parameter is determined by the parent space group $\mathcal G_0$, the parent irrep, and the chosen order parameter direction.
For the selected order parameter direction, the corresponding subgroup setting provides the basis transformation and, when applicable, the origin shift.
These relations define the supercell in the low symmetry setting and thereby determine how the relevant propagation vector is folded to the $\Gamma$ point when expressed in the reciprocal basis of $\mathcal G$.

To determine the real space representation of the N\'{e}el order in the isotropy subgroup, we subduce the parent representation to $\mathcal G$,
$\Gamma_N^{\mathcal G}=\Gamma_N(\mathbf q')\downarrow \mathcal G$,
with $\mathbf q'$ mapped to the folded $\Gamma$ point by the above supercell relation.
In practice, the subduction and the identification of the resulting irreps in $\mathcal G$ can be carried out using the representation tools hosted by the Bilbao Crystallographic Server~\cite{Aroyo:xo5013}, which implement the required correlation and compatibility analysis between the parent group and the subgroup.

\section{Ligand-rotation tight-binding models}\label{sec:2}
In this section, we provide a detailed derivation of the tight-binding (TB) models for the $d$-, $g$-, and $i$-wave antiferroaxial altermagnets introduced in the main text. These models explicitly incorporate the ligand degrees of freedom to elucidate the microscopic origin of the nonrelativistic spin splitting. The axial-vector IR within the spin-rotation space, denoted $\Gamma_A^s$, are omitted here and throughout the subsequent analysis.

\begin{figure*}[t]
\centering
\includegraphics[width=0.90\textwidth]{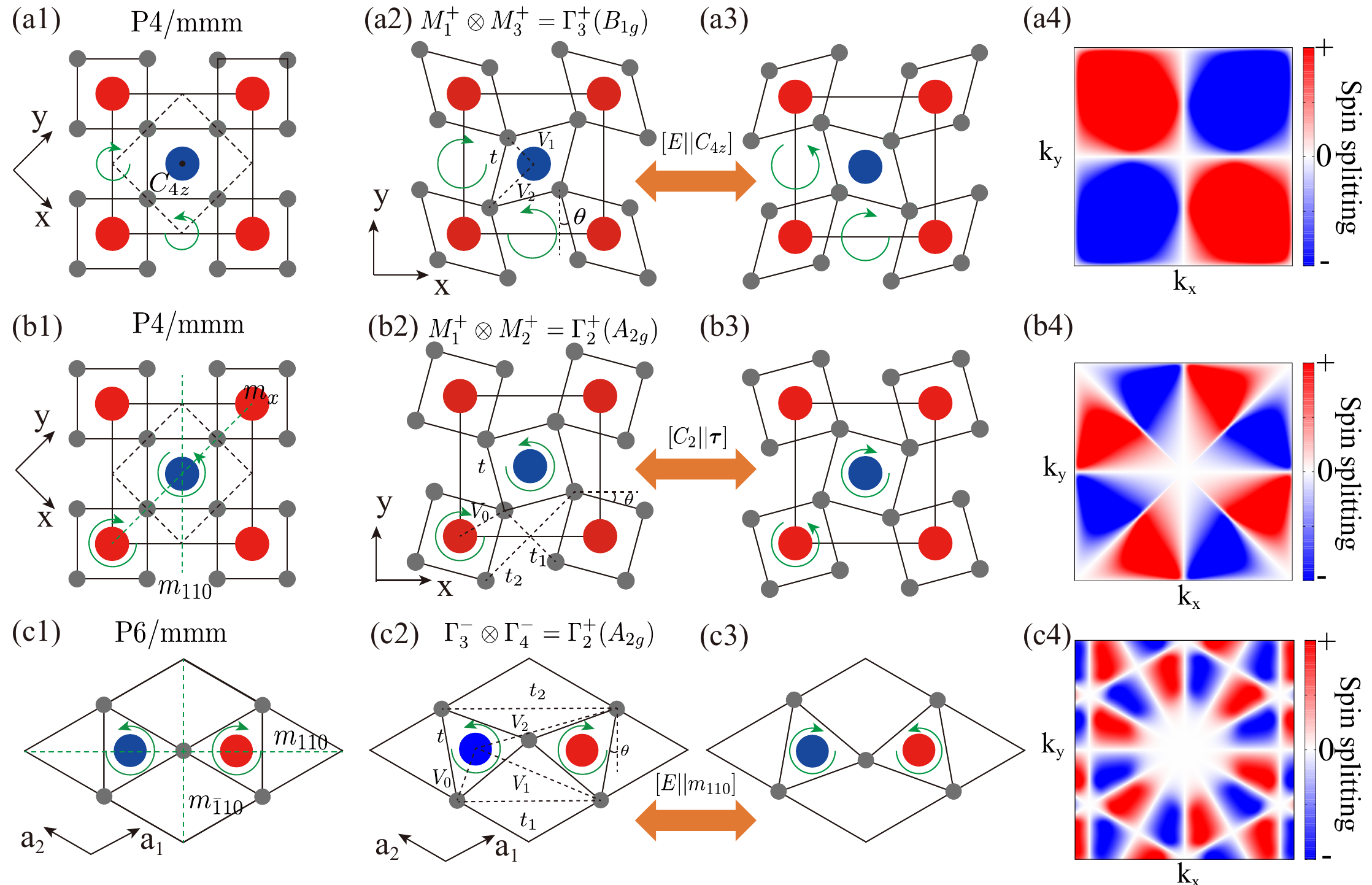}
\caption{\label{fig:TB}
Ligand-rotation tight-binding models for antiferroaxial altermagnets with $d$-, planar $g$-, and planar $i$-wave altermagnetic spin splitting.
Red (blue) spheres denote magnetic sites with spin up (down), gray spheres denote nonmagnetic ligand sites, and green curved arrows indicate the antiferroaxial rotation pattern parametrized by the ligand rotation angle $\theta$.
Dashed bonds mark the symmetry-inequivalent hopping/hybridization paths used in Eq.~(\ref{eq:hamiltonian}).
(a1)-(a4) $d$-wave model.
(a1) Prototypic parent structure with space group $P4/mmm$.
(a2) Structure in the antiferroaxial phase; the coexistence of the N\'{e}el order $M_1^+\otimes\Gamma_A^s$ and the antiferroaxial order $M_3^+$ induces the $\Gamma_3^+(B_{1g})$ multipole channel (corresponding to $d_{xy}$-type spin splitting).
(a3) Configuration with reversed antiferroaxial order ($\theta\to-\theta$), symmetry-equivalent to applying $[E|| C_{4z}]$ to (a2).
(a4) Resulting momentum-space spin splitting for (a3), showing the sign reversal relative to (a2), whose spin splitting is plotted in Fig.2 (a3) in the main text.
(b1)-(b4) planar $g$-wave model.
(b1) Parent phase ($P4/mmm$); green dashed lines indicate the mirror planes $m_x$ and $m_{110}$.
(b2) Antiferroaxial structure and the corresponding hopping network; the bilinear $M_1^+\otimes M_2^+$ yields the $\Gamma_2^+(A_{2g})$ sector, giving a planar $g$-wave splitting.
(b3) Antiferroaxial reversal ($\theta\to -\theta$), symmetry-equivalent to $[C_2||\boldsymbol{\tau}]$ applied to (b2) with $\boldsymbol{\tau}=(1/2,1/2,0)$.
(b4) Spin splitting for (b3), with the opposite sign compared to (b2), whose spin splitting is plotted in Fig.2 (b3) in the main text.
(c1)-(c4) planar $i$-wave model.
(c1) Parent structure with $P6/mmm$ symmetry; green dashed lines indicate the mirror planes $m_{110}$ and $m_{\bar{1}10}$.
(c2) Antiferroaxial structure and hopping network; the coexistence of $\mathbf N\sim \Gamma_3^-\otimes\Gamma_A^s$ and $\mathbf G\sim \Gamma_4^-$ induces the $\Gamma_2^+(A_{2g})$ sector, corresponding to planar $i$-wave splitting.
(c3) Reversed rotation pattern ($\theta\to-\theta$), symmetry-equivalent to applying $[E|| m_{\bar{1}10}]$ to (c2).
(c4) Spin splitting for (c3), exhibiting the corresponding sign reversal, whose spin splitting is plotted in Fig.2 (c3) in the main text.
The common model parameters are $J=t$ and $t_M=0.5t$, with specific values: (a) $\theta=20^\circ$, $V_1^0=V_2^0=0.6 t$. (b) $\theta=20^\circ$, $V_0=1.2 t$, $t_1^0=t_2^0=0.25 t$. (c) $\theta=30^\circ$, $V_0=1.2 t$, $V_1^0=V_2^0=0.5 t$, $t_1^0=t_2^0=0.2 t$.}
\end{figure*}

The TB Hamiltonian can be written as
\begin{equation}
\mathcal{H} =t_M\sum_{\langle mn\rangle} d_m^\dagger d_n
+ \sum_{mi}\Big[ V_{mi}(\theta)\, d_m^\dagger c_i + \mathrm{h.c.} \Big] + \sum_{ij} t_{ij}(\theta)\, c_i^\dagger c_j
+ J\sum_m \eta_m\, d_m^\dagger \sigma_z d_m ,
\label{eq:hamiltonian}
\end{equation}
where $d_m^\dagger$ ($c_i^\dagger$) creates an electron on the magnetic (ligand) site, and $\eta_m=\pm 1$ encodes the staggered N\'{e}el order.
The first term describes the direct nearest-neighbor hopping between magnetic sites with amplitude $t_M$.
The second and third terms describe the magnetic-ligand hybridization $V_{mi}(\theta)$ and ligand-ligand hopping $t_{ij}(\theta)$, respectively, both controlled by the antiferroaxial rotation angle $\theta$ through the bond-length dependence of hopping integrals [e.g., $V_{1,2}$ and $t_{1,2}$ in Fig.~\ref{fig:TB}(a2)-(c2)].
Following the main text, we incorporate this structural dependence via Harrison's scaling law $t(r)\propto r(\theta)^{-2}$~\cite{PhysRevB.20.2420}.
The last term introduces a local exchange field of strength $J$, with $\sigma_z$ the Pauli matrix and the quantization axis chosen along the collinear moments.

\subsection{$d$-wave antiferroaxial altermagnetic tight-binding model}

We construct the $d$-wave model based on the parent space group $P4/mmm$, as shown in Fig.~\ref{fig:TB}(a1). Both the antiferromagnetic and antiferroaxial instabilities occur at the $M$ point with wavevector $\mathbf{q}=(\pi,\pi,0)$, transforming as $\mathbf{N}\sim M_1^+$ and $\mathbf{G}\sim M_4^+\otimes(\Gamma_2^+\oplus\Gamma_5^+)$, respectively. Here, $M_4^+$ is the representation of the staggered pattern for the antiferroaxial order, which implies that the clockwise and counterclockwise rotational units are exchanged under $C_{4z}$ and $m_x$, but remain invariant under $m_{110}$ [the operators are plotted in Fig.~\ref{fig:TB}(b1)]. For the N\'{e}el vector, the spin-rotation space representation $\Gamma_A^s$ is suppressed, as the condition $\Gamma_1^s\subset \Gamma_A^s\otimes\Gamma_A^s$ is always satisfied. The ferroaxial order $\Gamma_A$ in the point group $4/mmm$ (corresponding to the space group $P4/mmm$) decomposes into the irreducible representation $\Gamma_2^+ \oplus \Gamma_5^+$. For the case with $\mathbf{G}\parallel \hat{z}$, we take $G_z \sim M_4^+\otimes\Gamma_2^+ = M_3^+$. The bilinear combination then transforms as
\begin{equation}
\mathbf{N}G_z\sim M_1^+\otimes M_3^+=\Gamma_3^+(B_{1g}),
\end{equation}
which enables a symmetry-allowed trilinear invariant with the rank-2 magnetic multipole sector in the parent group. Upon condensation of $G_z$, the symmetry is reduced to $P4/mbm$ with crystallographic point group $4/mmm$. In this subgroup, the N\'{e}el vector, which transforms as $M_1^+$ in the parent phase, subduces to the one-dimensional irrep $B_{2g}$. By consulting Table~\ref{table:altermagnet_valid_only}, we identify $B_{2g}$ with a $d_{xy}$-wave altermagnetic spin splitting and the corresponding spin point group $^{\bar{1}}4/m^{\bar{1}}mm$.

In the TB model, we include nearest-neighbor hopping between magnetic atoms ($t_M$) and between ligand atoms ($t$). For magnetic-ligand hybridization, we include both nearest- and next-nearest-neighbor hoppings ($V_1$ and $V_2$), as illustrated in Fig.~\ref{fig:TB}(a2), which can be modulated by the antiferroaxial distortion angle $\theta$ through Harrison's rule~\cite{PhysRevB.20.2420}
\begin{equation}
\begin{split}
V_1(\theta) &= V_1^0\frac{|\mathbf{r}(0)|^2}{|\mathbf{r}(\theta)|^2}
= V_1^0\frac{\sin^2(\pi/4)}{\sin^2(\pi/4-\theta)}
= \frac{V_1^0}{1-\sin(2\theta)},\\
V_2(\theta) &= V_2^0\frac{|\mathbf{r}(0)|^2}{|\mathbf{r}(\theta)|^2}
= V_2^0\frac{\sin^2(\pi/4)}{\sin^2(\pi/4+\theta)}
= \frac{V_2^0}{1+\sin(2\theta)},
\end{split}
\end{equation}
where we assume the ligand-ligand bond length remains constant so that $t$ is independent of $\theta$. The factor $\sin(\pi/4)$ and $\sin(\pi/4\pm\theta)$ are the hopping distance parameters before and after the distortion, respectively.

Reversing the antiferroaxial order, $\theta\to-\theta$, yields the structure in Fig.~\ref{fig:TB}(a3).
This configuration is symmetry-equivalent to applying $[E||C_{4z}]$ to Fig.~\ref{fig:TB}(a2), where $E$ denotes the identity in spin space and $C_{4z}$ represents a four-fold rotation in real space.
As a result, the spin splitting reverses sign, consistent with the reversal of the corresponding magnetic multipole component (cf.~Fig.~2(a3) in the main text).
\subsection{$g$-wave antiferroaxial altermagnetic tight-binding model}
For the planar $g$-wave antiferroaxial altermagnetic model, we consider the tetragonal parent space group $P4/mmm$, as illustrated in Fig.~\ref{fig:TB}(b1), where green dashed lines indicate the $m_x$ and $m_{110}$ mirror planes. In contrast to the $d$-wave case, the antiferromagnetic and antiferroaxial instabilities here share the same spatial modulation at the $M$ point, i.e., $\Gamma_N(\mathbf{q}')=\Gamma_G(\mathbf{q})=M_1^+$ with $\mathbf{q}=\mathbf{q}'=(\pi,\pi,0)$. Consequently, we have $G_z\sim M_1^+\otimes\Gamma_2^+=M_2^+$ and $\mathbf{N}\sim M_1^+$. The bilinear combination transforms as
\begin{equation}
\mathbf{N} G_z \sim M_1^+\otimes M_2^+ = \Gamma_2^+(A_{2g}),
\end{equation}
which allows coupling to the $A_{2g}$ channel of the rank-4 magnetic multipole $\mathbf{O}^{(4)}$ (see Table~\ref{table:TB1}). Upon condensation of $G_z$, the symmetry is lowered to the isotropy subgroup $P4/mbm$; notably, the crystallographic point group remains $4/mmm$. In this subgroup, the N\'{e}el order parameter (originally $M_1^+$) subduces to the $A_{2g}$ irreducible representation. As listed in Table~\ref{table:altermagnet_valid_only}, this representation corresponds to a planar $g$-wave altermagnetic spin splitting characterized by the spin point group $4/m^{\bar{1}}m^{\bar{1}}m$.

The Hamiltonian follows Eq.~(\ref{eq:hamiltonian}), with hopping paths shown in Fig.~\ref{fig:TB}(b2).
The antiferroaxial rotation angle $\theta$ modulates the ligand-ligand hoppings $t_1$ and $t_2$, while the base hoppings' amplitude $t$ and $V_0$ remain invariant. Their functional forms are given by
\begin{equation}
\begin{split}
t_1(\theta) &= t_1^0
\frac{\sin^2(\pi/4)}{\sin^2(\pi/4-\theta)}
= \frac{t_1^0}{1-\sin(2\theta)},\\
t_2(\theta) &= t_2^0
\frac{\sin^2(\pi/4)}{\sin^2(\pi/4+\theta)}
= \frac{t_2^0}{1+\sin(2\theta)},
\end{split}
\end{equation}
where the factors $\sin(\pi/4)$ and $\sin(\pi/4\pm\theta)$ in the numerators correspond to the hopping distance parameters before and after the distortion, respectively.

Reversing the antiferroaxial order, $\theta\to-\theta$, gives the structure in Fig.~\ref{fig:TB}(b3).
This configuration is symmetry-equivalent to applying $[C_2||\boldsymbol{\tau}]$ to Fig.~\ref{fig:TB}(b2), where $C_2$ denotes a $\pi$ rotation in spin space about an axis perpendicular to the ordered moments and $\boldsymbol{\tau}=(1/2,1/2,0)$ is a real-space translation.
As a result, the spin splitting reverses sign, as illustrated in Fig.~\ref{fig:TB}(b4) (cf.~Fig.~2(b3) in the main text).

\subsection{$i$-wave antiferroaxial altermagnetic tight-binding model}

For the planar $i$-wave antiferroaxial altermagnetic model, we adopt the hexagonal parent space group $P6/mmm$, as illustrated in Fig.~\ref{fig:TB}(c1), where the green dashed lines indicate the $m_{110}$ and $m_{\bar{1}10}$ mirror planes. In this hexagonal geometry, both the antiferromagnetic and antiferroaxial instabilities occur at the $\Gamma$ point. The staggered spatial modulation, which is shared by the N\'{e}el and antiferroaxial patterns, transforms as the one-dimensional representation $\Gamma_3^-$. Consequently, the N\'{e}el vector transforms as $\mathbf{N}\sim \Gamma_3^-$. The antiferroaxial order parameter $G_z$ is constructed by combining this staggered modulation with the axial-vector representation $\Gamma_2^+(A_{2g})$ of the $6/mmm$ point group, yielding $G_z \sim \Gamma_3^- \otimes \Gamma_2^+ = \Gamma_4^-$. The resulting bilinear combination transforms as
\begin{equation}
\mathbf{N}G_z \sim \Gamma_3^- \otimes \Gamma_4^- = \Gamma_2^+(A_{2g}).
\end{equation}
This coupling enables a symmetry-allowed trilinear invariant with the $A_{2g}$ channel of the rank-6 magnetic multipole $\mathbf{O}^{(6)}$ (see Table~\ref{table:TB1}). Upon condensation of $G_z$, the symmetry is reduced to the isotropy subgroup $P\bar{6}m2$. In this subgroup, the N\'{e}el vector (originally $\Gamma_3^-$) subduces to the representation $A_2'$. As listed in Table~\ref{table:altermagnet_valid_only}, this representation corresponds to a planar $i$-wave altermagnetic spin splitting.

The tight-binding Hamiltonian includes the nearest-neighbor hopping between magnetic atoms ($t_M$), ligand-ligand hopping ($t$), and magnetic-ligand hybridization ($V$). The antiferroaxial rotation $\theta$ lifts the symmetry of the ligand environment, resulting in two inequivalent ligand-ligand hoppings ($t_1, t_2$) and two inequivalent magnetic-ligand hybridizations ($V_1, V_2$), as shown in Fig.~\ref{fig:TB}(c2). Incorporating the structural dependence via Harrison's scaling law, the hopping amplitudes are given by
\begin{equation}
\begin{split}
t_{1}(\theta) &= t_{1}^0\frac{(\sqrt{3})^2}{2-2\cos(2\pi/3+ 2\theta)}= \frac{3t_{1}^0}{2\left[1+\cos(2\theta - \pi/3)\right]},\\
t_{2}(\theta) &= t_{2}^0\frac{(\sqrt{3})^2}{2-2\cos(2\pi/3- 2\theta)}= \frac{3t_{2}^0}{2\left[1+\cos(2\theta + \pi/3)\right]},\\
V_{1}(\theta) &= V_{1}^0\frac{7/3}{4/3\cos^2\theta+1/3-4/3\cos\theta\cos(2\pi/3+\theta)} =\frac{7V_{1}^0}{4 + 2\sqrt{3}\sin(2\theta + \pi/3)},\\
V_{2}(\theta) &= V_{2}^0\frac{7/3}{4/3\cos^2\theta+1/3-4/3\cos\theta\cos(2\pi/3-\theta)} =\frac{7V_{2}^0}{4 - 2\sqrt{3}\sin(2\theta - \pi/3)},
\end{split}
\end{equation}
where the denominators reflect the changes in bond lengths induced by the distortion angle $\theta$.

Reversing the antiferroaxial order, $\theta\to-\theta$, yields the structure in Fig.~\ref{fig:TB}(c3).
This configuration is related to Fig.~\ref{fig:TB}(c2) by the symmetry operation $[E\Vert m_{\bar{1}10}]$, with $E$ the identity in spin space.
Accordingly, the induced $\Gamma_2^+(A_{2g})$ multipole and the associated planar $i$-wave altermagnetic spin splitting change sign under $\theta\to-\theta$, consistent with the reversal shown in the main text [Fig.~2(c3)].

\section{Landau Theory}\label{sec:3}
\label{sec:Landau}

In this section, we derive an effective Landau functional for cubic perovskites starting from the parent space group $Pm\bar{3}m$, with an emphasis on the $R$-point structural instability at $\mathbf{q}=(\pi,\pi,\pi)$. This high-symmetry parent phase is shared by a broad class of antiferroaxial altermagnetic materials, including perovskite oxides (e.g., KMnF$_3$, LaCrO$_3$) and transition-metal trifluorides (e.g., FeF$_3$).

\subsection{Order parameters and symmetry}
We introduce the following order parameters, classified by the irreducible representations (IRs) of the parent space group at the $R$ point:

\begin{enumerate}
\item \textbf{N\'{e}el order.} The antiferromagnetic order parameter is denoted by the spin-space vector $\mathbf N$ and transforms as
$\mathbf N\sim R_1^+$.
Throughout this section, we take the zero SOC limit, the direction of $\mathbf N$ is decoupled from the real space, and we treat its amplitude $N\equiv|\mathbf N|$ as the relevant Landau variable.

\item \textbf{Antiferroaxial order.} The antiferroaxial (counter-rotating) distortion is described by a three-component axial order parameter
$\mathbf G=(G_x,G_y,G_z)$ transforming as $\mathbf G\sim R_1^+\otimes\Gamma_4^+(T_{1g})= R_4^+$.

\item \textbf{Elastic strains.}
Since $\mathbf G$ is a structural order parameter, the lattice strains $\varepsilon_{ij}$ constitute additional symmetry-allowed secondary degrees of freedom and generally couple to $\mathbf G$ via rotostriction, thereby renormalizing the effective Landau coefficients and generating anisotropy.
We therefore include the symmetric strain tensor $\varepsilon_{ij}$ and decompose it into symmetry-adapted combinations of the cubic point group $O_h$:
\begin{align}
& A_{1g}: && \varepsilon_v=\varepsilon_{xx}+\varepsilon_{yy}+\varepsilon_{zz}, \\
& E_g: && \varepsilon_1=\varepsilon_{zz}-\tfrac{1}{2}(\varepsilon_{xx}+\varepsilon_{yy}),\qquad
\varepsilon_2=\tfrac{\sqrt{3}}{2}(\varepsilon_{yy}-\varepsilon_{xx}),\\
& T_{2g}: && (\varepsilon_{yz},\varepsilon_{xz},\varepsilon_{xy}).
\end{align}

\item \textbf{Rank-4 (g-wave) magnetic multipoles.}
The antiferroaxial altermagnetic order is diagnosed by the rank $\ell=4$ magnetic multipole in the $T_{1g}$ spatial channel.
We choose a convenient basis
\begin{equation}
(\mathbf O_1,\mathbf O_2,\mathbf O_3)
=(\mathbf O_{yz(y^2-z^2)},\ \mathbf O_{xz(x^2-z^2)},\ \mathbf O_{xy(x^2-y^2)}),
\end{equation}
where each $\mathbf O_i$ is a spin-space vector defined microscopically by
$\mathbf O_{xy(x^2-y^2)}\equiv \int d^3r\, xy(x^2-y^2)\,\mathbf m(\mathbf r)$,
and similarly for the other components.
\end{enumerate}
Because $\mathbf N$ and $\mathbf O_i$ are both vectors in spin space, all dot products below are taken in spin space.

\subsection{Landau free energy}
The total free-energy density, which can be constructed using the symmetry analysis tools in the ISOTROPY software suite~\cite{stokes2016isosubgroup,hatch2003invariants}, is given by
\begin{equation}
\mathcal F = \mathcal F_{N,G}+\mathcal F_{\varepsilon}+\mathcal F_{G\varepsilon}+\mathcal F_{N,G,O}.
\end{equation}

We first collect the magnetic and antiferroaxial contributions. Keeping terms up to quartic order, we write
\begin{align}
\mathcal F_{N,G}
=&\ \alpha_N \mathbf N^2 + \beta_N (\mathbf N^2)^2
+ \alpha_G \mathbf G^2 + \beta_{G1}(\mathbf G^2)^2
+ \beta_{G2}\!\left(G_x^2G_y^2+G_x^2G_z^2+G_y^2G_z^2\right)
+ \lambda_{NG}\mathbf N^2\mathbf G^2,
\label{eq:SM_F_NG}
\end{align}
where $\mathbf G^2=G_x^2+G_y^2+G_z^2$.

The elastic energy at the harmonic level allowed by $O_h$ is
\begin{equation}
\mathcal F_{\varepsilon}
=C_0\varepsilon_v^2
+C_1(\varepsilon_1^2+\varepsilon_2^2)
+C_2(\varepsilon_{yz}^2+\varepsilon_{xz}^2+\varepsilon_{xy}^2).
\label{eq:SM_F_elastic}
\end{equation}

The bilinear coupling between the quadratic form of $\mathbf G$ and the strain irreducible representations is
\begin{align}
\mathcal F_{G\varepsilon}
=&\ \lambda_1\!\left[
\frac{1}{\sqrt{3}}\left(2G_z^2-G_x^2-G_y^2\right)\varepsilon_1
+\left(G_x^2-G_y^2\right)\varepsilon_2
\right]
+\lambda_2\left(G_yG_z\,\varepsilon_{yz}+G_xG_z\,\varepsilon_{xz}+G_xG_y\,\varepsilon_{xy}\right).
\label{eq:SM_F_rotostriction}
\end{align}
An additional $A_{1g}$ coupling $\propto \mathbf G^2\varepsilon_v$ is symmetry-allowed and can be absorbed into a redefinition of $\beta_{G1}$.

Finally, the lowest-order invariant coupling among $\mathbf N$, $\mathbf G$, and the $T_{1g}$ multipoles reads
\begin{equation}
\mathcal F_{N,G,O}
=\lambda_O\left(\mathbf O_1^2+\mathbf O_2^2+\mathbf O_3^2\right)
+g\,\mathbf N\cdot\left(G_x\mathbf O_1+G_y\mathbf O_2+G_z\mathbf O_3\right).
\label{eq:SM_F_NGO}
\end{equation}
Once both $\mathbf N\neq 0$ and $\mathbf G\neq 0$ develop, this term induces the $g$-wave multipole as $\mathbf O_i\propto \mathbf N G_i$, implying an antiferroaxial control of the altermagnetic spin splitting.

\subsection{Integrating out secondary degrees of freedom}
We now derive an effective free energy in terms of the primary structural order parameter $\mathbf G$ by eliminating
$\mathbf O$, $\mathbf N$ (amplitude), and the strains $\varepsilon$ through their saddle-point conditions.

\subsubsection{Eliminating the multipoles $\mathbf O$}
Minimizing Eq.~\eqref{eq:SM_F_NGO} with respect to $\mathbf O_i$ gives
\begin{equation}
\frac{\partial \mathcal F}{\partial \mathbf O_i}=0
\quad\Rightarrow\quad
\mathbf O_i=-\frac{g}{2\lambda_O}\,\mathbf N\,G_i.
\label{eq:SM_O_constitutive}
\end{equation}
Substituting back into Eq.~\eqref{eq:SM_F_NGO} yields a negative biquadratic contribution
\begin{equation}
\mathcal F_{N,G,O}\ \to\ -\frac{g^2}{4\lambda_O}\,\mathbf N^2\mathbf G^2.
\end{equation}
Therefore the effective $\mathbf N^2\mathbf G^2$ coupling becomes
\begin{equation}
\Lambda \equiv \lambda_{NG}-\frac{g^2}{4\lambda_O}.
\label{eq:SM_Lambda_def}
\end{equation}

\subsubsection{Eliminating the N\'{e}el amplitude $N$}
After eliminating $\mathbf O$, the $N$-dependent part is
\begin{equation}
\mathcal F_N
=\alpha_N \mathbf N^2+\beta_N(\mathbf N^2)^2+\Lambda\,\mathbf N^2\mathbf G^2.
\end{equation}
Stationarity $\partial\mathcal F/\partial \mathbf N=0$ gives
\begin{equation}
\alpha_N\mathbf N + 2\beta_N\mathbf N^2\mathbf N +\Lambda\,\mathbf G^2\mathbf N=0.
\end{equation}
In the magnetically ordered phase ($\mathbf N\neq 0$) this yields
\begin{equation}
\mathbf N^2=-\frac{\alpha_N+\Lambda\,\mathbf G^2}{2\beta_N}.
\label{eq:SM_N2}
\end{equation}
Substituting Eq.~\eqref{eq:SM_N2} back contributes the standard negative-definite term
\begin{equation}
\mathcal F_N \ \to\ -\frac{(\alpha_N+\Lambda\,\mathbf G^2)^2}{4\beta_N}
=\mathrm{const.}-\frac{\alpha_N\Lambda}{2\beta_N}\mathbf G^2-\frac{\Lambda^2}{4\beta_N}(\mathbf G^2)^2.
\end{equation}

\subsubsection{Eliminating strains}
Minimizing $\mathcal F_\varepsilon+\mathcal F_{G\varepsilon}$ with respect to each strain component gives
\begin{equation}
\begin{split}
\varepsilon_1&=-\frac{\lambda_1}{2C_1}\frac{1}{\sqrt{3}}\left(2G_z^2-G_x^2-G_y^2\right),\qquad
\varepsilon_2=-\frac{\lambda_1}{2C_1}\left(G_x^2-G_y^2\right), \\
\varepsilon_{ij}&=-\frac{\lambda_2}{2C_2}G_iG_j,\qquad (i\neq j;\ i,j\in\{x,y,z\}),
\end{split}
\label{eq:SM_strain_eq}
\end{equation}
while $\varepsilon_v$ decouples in the present truncation.
Substituting \eqref{eq:SM_strain_eq} back yields the strain-mediated quartic renormalization
\begin{equation}
\Delta \mathcal F_{\rm strain}
=-\frac{\lambda_1^2}{3C_1}(\mathbf G^2)^2+\left(\frac{\lambda_1^2}{C_1}-\frac{\lambda_2^2}{4C_2}\right)\left(G_x^2G_y^2+G_x^2G_z^2+G_y^2G_z^2\right).
\end{equation}

\subsection{Effective free energy for $\mathbf G$}
Collecting the above results, the effective free energy takes the cubic-invariant form
\begin{equation}
\mathcal F_{\rm eff}(\mathbf G)
=\alpha_{\rm eff}\mathbf G^2
+\beta_{1,{\rm eff}}(\mathbf G^2)^2
+\beta_{2,{\rm eff}}\left(G_x^2G_y^2+G_x^2G_z^2+G_y^2G_z^2\right),
\label{eq:SM_Feff}
\end{equation}
with renormalized coefficients
\begin{align}
\alpha_{\rm eff}&=\alpha_G-\frac{\alpha_N}{2\beta_N}\Lambda,\\
\beta_{1,{\rm eff}}&=\beta_{G1}-\frac{\Lambda^2}{4\beta_N}-\frac{\lambda_1^2}{3C_1},\\
\beta_{2,{\rm eff}}&=\beta_{G2}+\frac{\lambda_1^2}{C_1}-\frac{\lambda_2^2}{4C_2}.
\label{eq:SM_coeffs}
\end{align}
The anisotropy parameter $\beta_{2,{\rm eff}}$ determines the orientation of $\mathbf G$ and therefore the form of the induced $g$-wave multipole.

\subsection{Phase I: $\mathbf G \parallel [001]$ (planar $g$-wave)}
For $\beta_{2,{\rm eff}}>0$, the minimum is achieved when $\mathbf G$ points along a principal axis, e.g.
$\mathbf G=(0,0,G_z)$.
Then Eq.~\eqref{eq:SM_Feff} reduces to
\begin{equation}
\mathcal F_{\rm eff}=\alpha_{\rm eff}G_z^2+\beta_{1,{\rm eff}}G_z^4,
\end{equation}
with $G_z^2=-\alpha_{\rm eff}/(2\beta_{1,{\rm eff}})$ in the ordered phase.
Stability against transverse fluctuations requires $\beta_{1,{\rm eff}}>0$ and $\beta_{2,{\rm eff}}>0$.

The secondary orders follow from Eq.~\eqref{eq:SM_O_constitutive}, Eq.~\eqref{eq:SM_N2}, and Eq.~\eqref{eq:SM_strain_eq}:
\begin{equation}
\mathbf O_3=-\frac{g}{2\lambda_O}\mathbf N G_z\neq 0,\quad
\mathbf O_1=\mathbf O_2=0,\quad \mathbf{N}^2=-\frac{\alpha_N+\Lambda G_z^2}{2\beta_N},
\end{equation}
indicating a planar $g$-wave multipole with $xy(x^2-y^2)$ spin-splitting form.
The equilibrium strains are
\begin{equation}
\varepsilon_2=\varepsilon_{xy}=\varepsilon_{xz}=\varepsilon_{yz}=0,\qquad
\varepsilon_1=-\frac{\lambda_1}{C_1\sqrt{3}}G_z^2.
\end{equation}

\textit{When coupling to uniaxial stress along $[001]$.}
Applying a uniaxial stress $\sigma_z$ adds
$\mathcal F_{\rm stress}=-\sigma_z\varepsilon_{zz}$.
Using $\varepsilon_{zz}=\varepsilon_v/3+2\varepsilon_1/3$, only $\varepsilon_1$ is relevant here.
The minimization condition becomes
\begin{equation}
2C_1\varepsilon_1+\frac{\lambda_1}{\sqrt{3}}(2G_z^2)-\frac{2}{3}\sigma_z=0
\quad\Rightarrow\quad
\varepsilon_1=-\frac{\lambda_1}{C_1\sqrt{3}}G_z^2+\frac{\sigma_z}{3C_1}.
\end{equation}
Substituting this solution back, the stress generates a linear shift of the quadratic coefficient,
\begin{equation}
\alpha_{\rm eff}(\sigma_z)=\alpha_{\rm eff}(0)+\frac{2\lambda_1}{3\sqrt{3}C_1}\sigma_z.
\end{equation}
This leads to the relation $G_z \propto \sqrt{\sigma_z - \sigma_c}$, valid for $\lambda_1 \sigma_z > 0$, where the critical threshold is given by $\sigma_c = -\frac{3\sqrt{3}C_1}{2\lambda_1}\alpha_{\rm eff}(0)$. This result demonstrates that the antiferroaxial order---and consequently the altermagnetism---can be mechanically controlled on or off. This scenario is realized in KMnF$_3$ [Fig.~\ref{fig:KMnF3}(d)].

\subsection{Phase II: $\mathbf G \parallel [111]$ (bulk $g$-wave)}
For $\beta_{2,{\rm eff}}<0$, the minimum is achieved for the body-diagonal configuration
$G_x=G_y=G_z=G/\sqrt{3}$,
yielding
\begin{equation}
\mathcal F_{\rm eff}=\alpha_{\rm eff}G^2+\beta_{\rm eff}G^4,\qquad
\beta_{\rm eff}\equiv \beta_{1,{\rm eff}}+\frac{1}{3}\beta_{2,{\rm eff}},
\end{equation}
and $G^2=-\alpha_{\rm eff}/(2\beta_{\rm eff})$.
The induced multipoles are equal-amplitude
\begin{equation}
\mathbf O_1=\mathbf O_2=\mathbf O_3=-\frac{g}{2\lambda_O}\mathbf N\frac{G}{\sqrt{3}},
\end{equation}
corresponding to the coherent superposition
$yz(y^2-z^2)+zx(z^2-x^2)+xy(x^2-y^2)=(x+y+z)(x-y)(y-z)(z-x)$, i.e., the bulk $g$-wave form.

In this phase the $E_g$-type strains vanish, hence $\varepsilon_1=\varepsilon_2=0$, while the $T_{2g}$ shear strains satisfy
\begin{equation}
\varepsilon_{xy}=\varepsilon_{yz}=\varepsilon_{xz}\equiv \varepsilon_s
=-\frac{\lambda_2}{6C_2}G^2.
\end{equation}

\begin{figure*}[htbp]
\includegraphics[width=0.6\textwidth]{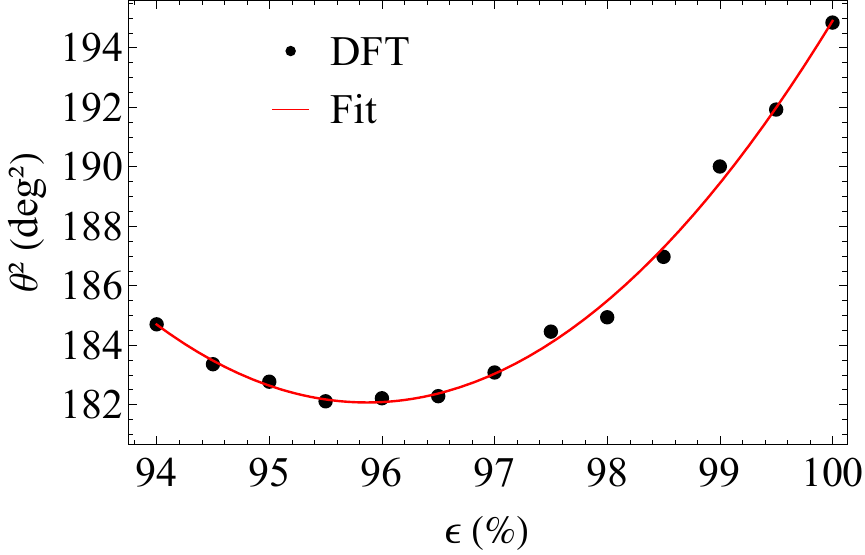}
\caption{\label{fig:strain} Dependence of the squared antiferroaxial rotation angle $\theta^2$ on the applied uniaxial strain $\varepsilon_{111}$ in FeF$_3$, obtained from DFT calculations. The numerical results are well fit to the quadratic relation $\theta^2=7085-144\varepsilon_{111}+0.7511\varepsilon_{111}^2$, where $\theta$ is expressed in degrees and the strain $\varepsilon_{111}$ is expressed in percentage.} \end{figure*}

\textit{Stress along $[111]$ and anharmonic elasticity.}
Under uniaxial stress $\sigma_{111}$, the relevant combination is
$\varepsilon_{111}=\frac{1}{3}\varepsilon_v+2\varepsilon_s$,
giving $\mathcal F_{\rm stress}=-(\sigma_{111}/3)\varepsilon_v-2\sigma_{111}\varepsilon_s$.
Because $\varepsilon_s\neq 0$, the lowest shear-anharmonic invariant
$C_3\varepsilon_{xy}\varepsilon_{yz}\varepsilon_{xz}=C_3\varepsilon_s^3$
contributes. The shear-sector free energy becomes
\begin{equation}
\mathcal F_s
=3C_2\varepsilon_s^2+C_3\varepsilon_s^3+\lambda_2G^2\varepsilon_s-2\sigma_{111}\varepsilon_s.
\end{equation}
Minimizing with respect to $\varepsilon_s$ gives
\begin{equation}
(\lambda_2G^2-2\sigma_{111})+6C_2\varepsilon_s+3C_3\varepsilon_s^2=0.
\end{equation}
For weak anharmonicity $|C_3|\ll C_2$, the solution satisfying $\varepsilon_s\to 0$ when $G,\sigma_{111}\to 0$ reads
\begin{equation}
\varepsilon_s(\sigma_{111})
\simeq \frac{2\sigma_{111}-\lambda_2G^2}{6C_2}
-\frac{C_3}{72C_2^3}\left(2\sigma_{111}-\lambda_2G^2\right)^2.
\end{equation}
Substituting back, the stress effectively renormalizes the quadratic and quartic coefficients (up to $O(G^4)$ and keeping the leading stress dependence):
\begin{align}
\alpha_{\rm eff}(\sigma_{111})
&=\alpha_{\rm eff}(0)+\frac{\lambda_2}{3C_2}\sigma_{111}
-\frac{C_3\lambda_2}{18C_2^3}\sigma_{111}^2,\\
\beta_{\rm eff}(\sigma_{111})
&=\beta_{\rm eff}(0)+\frac{C_3\lambda_2^2}{36C_2^3}\sigma_{111},
\end{align}
which implies a generically non-monotonic (parabolic) dependence of $G^2$ on $\sigma_{111}$ (or strain, as stress and strain satisfy Hooke's law in the small-strain limit).
This provides an analytic rationale for the parabolic strain dependence of the antiferroaxial rotation amplitude observed in FeF$_3$, as shown in Fig.~\ref{fig:strain}.

\section{Computational details and candidate antiferroaxial altermagnets}\label{sec:material}

\subsection{First-principles calculations and Wannier-based tight-binding models}
The first-principles calculations of structure optimization and band structures are using Vienna \textit{ab initio} simulation package(VASP)\cite{kresseEfficientIterativeSchemes1996}. The generalized gradient approximation (GGA) with Perdew-Burke-Ernzerhof (PBE)\cite{perdewGeneralizedGradientApproximation1996a} realization is considered. For correlated $3d$ orbitals, a Hubbard $U$ correction was employed as specified for each material below.

To evaluate Berry-curvature-related responses (e.g., anomalous Hall conductivity in FeF$_3$), we constructed \textit{ab initio} tight-binding Hamiltonians based on maximally localized Wannier functions using Wannier90 code~\cite{mostofiWannier90ToolObtaining2008}. Unless otherwise stated, the nonrelativistic altermagnetic spin splitting discussed in this work is analyzed in the zero SOC limit. In this section, we suppress the axial-vector representation $\Gamma_A^s$ in spin-rotational space.

In this section, we provide the computational details, symmetry analyses, and calculated physical properties for all candidate antiferroaxial altermagnets identified in this work, including MnS$_2$, MnC$_2$, FeBr$_3$, KMnF$_3$, Ca$_2$MnO$_4$, FeF$_3$, CoF$_3$, LaCrO$_3$, and La$_2$NiO$_4$.

\subsection{2D pentagonal planar $g$-wave altermagnets: MnA$_2$ (A = S, C)}
\label{subsec:MnA2}

\begin{figure*}[htbp]
\includegraphics[width=0.90\textwidth]{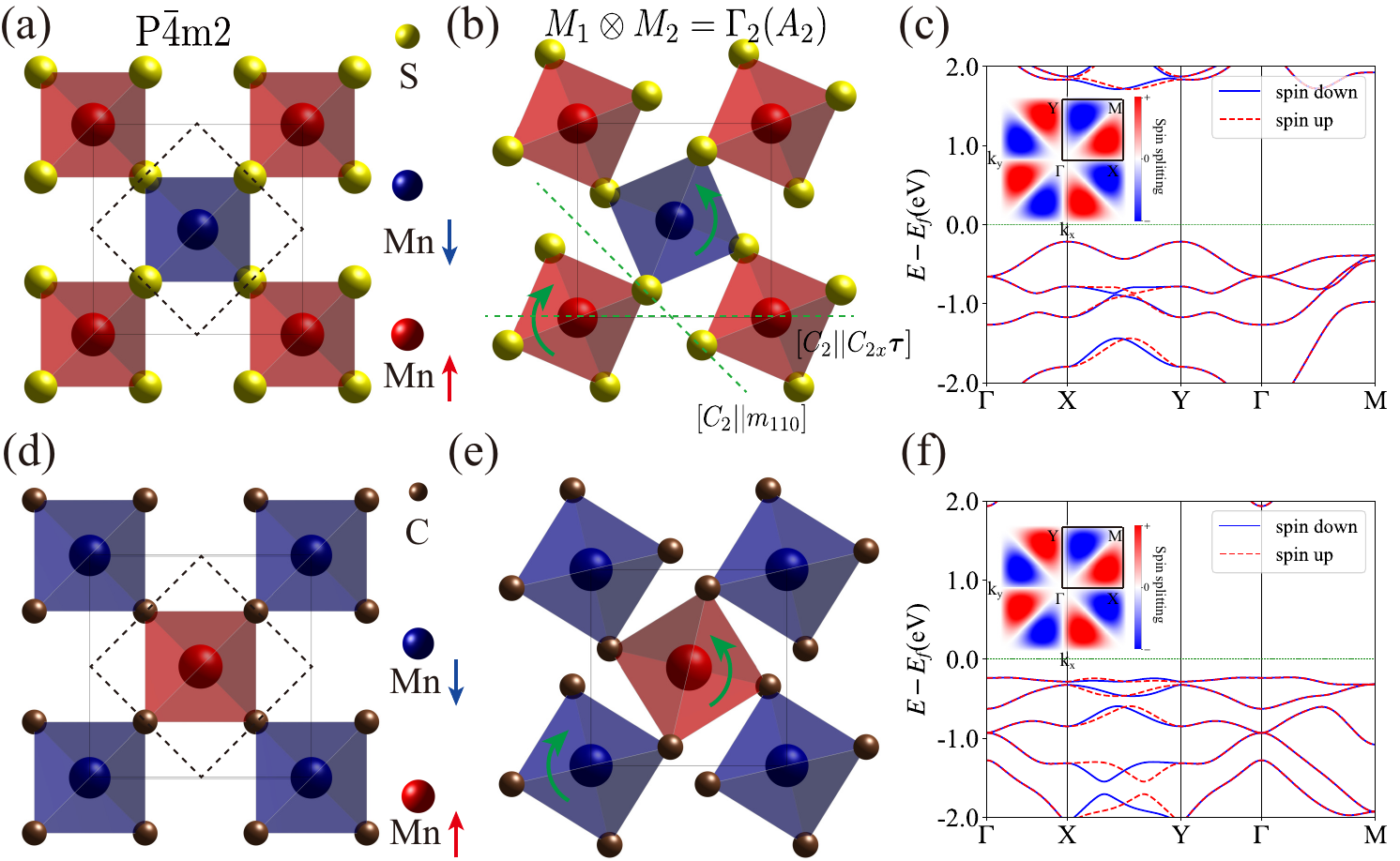}
\caption{\label{fig:MnS2}
Structural transition and electronic structure of MnA$_2$ (A=S, C).
(a,d) Parent phase (space group $P\bar{4}m2$). The primitive cell is shown in dashed lines. Red/blue Mn spheres denote opposite spins; yellow/brown spheres denote nonmagnetic ligands forming a tetrahedra. The N\'{e}el vector transforms as $M_1$.
(b,e) Antiferroaxial phase with counter-rotating ligand tetrahedra, described by $G_z\sim M_2$. After $G_z$ condensed, two opposite magnetic Mn atoms can be connected through spin group operator $[C_2||m_{110}]$ and $[C_2||C_{2x}\boldsymbol\tau]$, where $\boldsymbol\tau=(1/2,1/2,0)$.
(c,f) Band structures exhibiting planar $g$-wave altermagnetic spin splitting in the antiferroaxial phase (isotropy subgroup $P\bar{4}2_1m$), where the N\'{e}el order subduces to $\Gamma_2(A_2)$.}
\end{figure*}

Figure~\ref{fig:MnS2} summarizes the crystal structures, magnetic configurations, and electronic band structures of the 2D pentagonal systems MnS$_2$ and MnC$_2$~\cite{wangPentagonal2DAltermagnetsb}. The parent phase belongs to the space group $P\bar{4}m2$, where the primitive cell contains a magnetic Mn atom coordinated by nonmagnetic ligands, forming a M-A$_4$ (A=S, C) tetrahedral unit with $S_{4z}$ symmetry. Both the antiferromagnetic and antiferroaxial instabilities occur at the $M$ point $\mathbf{q}=(\pi,\pi,0)$, yielding a C-type antiferromagnetic order. In the zero SOC limit, the N\'{e}el vector transforms as $\mathbf{N}\sim M_1$, and the antiferroaxial order corresponds to the counter-rotation of neighboring ligand tetrahedra around the $z$ axis, characterized by a staggered axial vector $G_z$ transforming as $G_z\sim M_2$. Consequently, their bilinear product transforms as a $\Gamma$-point representation,
\begin{equation}
G_z\mathbf{N}\sim M_1\otimes M_2=\Gamma_2(A_2),
\end{equation}
and thus couples to the $A_2$ component of the rank four multipoles,
\begin{equation}
\mathbf{O}^{(4)}\sim 2A_1\oplus A_2\oplus B_1\oplus B_2\oplus 2E.
\end{equation}
The isotropy subgroup is $P\bar{4}2_1m$, in which the subduced N\'{e}el order transforms as $\Gamma_2(A_2)$, resulting in the planar $g$-wave altermagnetic spin splitting according to Table~\ref{table:altermagnet_valid_only}, consistent with DFT results shown in Fig.~\ref{fig:MnS2}(c,f).

\textit{Computational details:} For MnS$_2$ and MnC$_2$, we used $U=4$ eV on Mn $3d$ orbitals. The Brillouin zone was sampled using a $13\times 13\times 1$ k mesh, and the plane-wave cutoff energy was set to 550 eV.

\subsection{2D planar $i$-wave altermagnet: 2H-FeBr$_3$}
\label{subsec:FeBr3}

\begin{figure*}[htbp]
\includegraphics[width=0.70\textwidth]{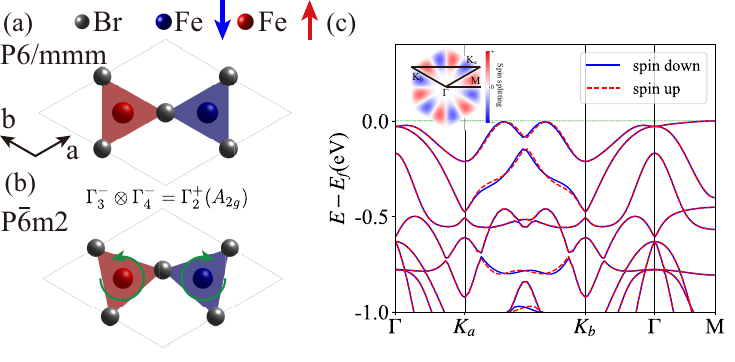}
\caption{\label{fig:FeBr3}
Antiferroaxial transition and planar $i$-wave altermagnetism in 2H-FeBr$_3$.
(a) Parent phase (space group $P6/mmm$) with Fe honeycomb and Br Kagome sublattices. The N\'{e}el vector transforms as $\Gamma_3^-$.
(b) Antiferroaxial phase: counter-rotation of Br ligands described by $G_z\sim \Gamma_4^-$, lowering the symmetry to $P\bar{6}m2$.
(c) Band structure exhibiting the planar $i$-wave altermagnetic spin splitting in the antiferroaxial phase.}
\end{figure*}

Figure~\ref{fig:FeBr3} shows the structure and band structure of 2H-FeBr$_3$~\cite{sodequistTwodimensionalAltermagnetsHigh2024c}. The parent phase belongs to $P6/mmm$, where in the parent phase Fe atoms form a honeycomb lattice and Br atoms form a Kagome network [Fig.~\ref{fig:FeBr3}(a)]. The two Fe sublattices (spin up/down) are related by spatial inversion in the parent phase, and the collinear N\'{e}el vector transforms as $\Gamma_3^-$ in the zero SOC limit. The antiferroaxial transition corresponds to the counter-rotation of Br ligands around the two inversion-related Fe sites, which transforms as $G_z \sim \Gamma_3^-\otimes \Gamma_2^+ = \Gamma_4^-$. Consequently, their bilinear product transforms as
\begin{equation}
G_z\mathbf{N}\sim\Gamma_3^-\otimes\Gamma_4^-=\Gamma_2^+(A_{2g}),
\end{equation}
and thus couples to the $A_{2g}$ component of the rank six multipoles,
\begin{equation}
\mathbf{O}^{(6)}\sim A_{1g}\oplus A_{2g}\oplus E_g\oplus T_{1g}\oplus 2T_{2g}.
\end{equation}
The isotropy subgroup is $P\bar{6}m2$, in which the subduced N\'{e}el order transforms as $\Gamma_2(A_2')$, resulting in the planar $i$-wave altermagnetic spin splitting according to Table~\ref{table:altermagnet_valid_only}, consistent with DFT results shown in Fig.~\ref{fig:FeBr3}(c)

\textit{Computational details:} We used $U=5$ eV on Fe $3d$ orbitals, a $13\times 13\times 1$ $k$ mesh, and a plane-wave cutoff energy of 550 eV.

\subsection{3D planar $g$-wave altermagnets: KMnF$_3$ and Ca$_2$MnO$_4$}
\label{subsec:KMnF3_Ca2MnO4}

\begin{figure*}[htbp]
\includegraphics[width=0.80\textwidth]{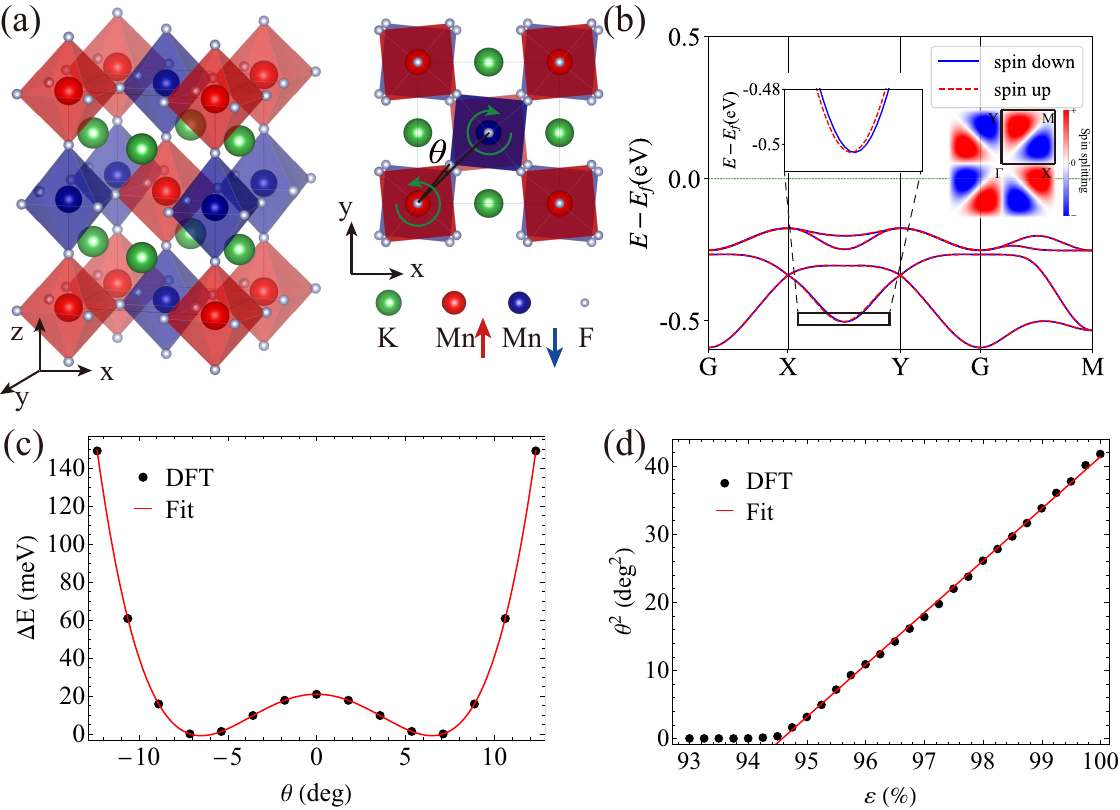}
\caption{\label{fig:KMnF3}
Structure, electronic properties, and energetics of the planar $g$-wave altermagnet KMnF$_3$.
(a) Crystal structure illustrating the staggered rotation of corner-sharing Mn-F$_6$ octahedra about the $z$ axis; red and blue octahedras denote the spin-up and spin-down sublattices, respectively.
(b) Electronic band structure exhibiting the characteristic planar $g$-wave altermagnetic spin splitting; the bands remain doubly degenerate along $\Gamma$-$X$, $\Gamma$-$Y$, and $\Gamma$-$M$.
(c) DFT total energy as a function of the octahedral rotation angle $\theta$ (black dots), fitted by an effective Landau potential (red line).
The fitted free energy per unit cell is
$\mathcal{F}_\mathrm{eff}(\theta)=20.8575-1.03319\,\theta^2+0.0122299\,\theta^4$,
where $\mathcal{F}_\mathrm{eff}$ is in meV per unit cell and $\theta$ is in degrees.
(d) Strain dependence of the structural order parameter: $\theta^2$ as a function of the uniaxial strain $\varepsilon_{zz}$.
Black dots are DFT data and the red line is a linear fit,
$\theta^2=-721.634+7.62945\varepsilon_{zz}$ (with $\theta$ in degrees and $\varepsilon_{zz}$ in percent). When $\varepsilon_{zz} < 94.5\%$, the angle $\theta=0$. Consequently, the spin splitting vanishes, and the system effectively reduces to a $T\tau$-invariant antiferromagnet.}
\end{figure*}

KMnF$_3$~\cite{knightNuclearMagneticStructures2020} is a representative perovskite-based planar $g$-wave altermagnet. The antiferroaxial order is realized by a staggered rotation of corner-sharing MnF$_6$ octahedra, which we parameterize by the rotation angle $\theta$, and the magnetic ground state is the $G$-type antiferromagnet [Fig.~\ref{fig:KMnF3}(a)]. Within the Landau description in Sec.~\ref{sec:3}, the planar character is selected when the antiferroaxial anisotropy favors a principal-axis configuration, e.g., $\mathbf G \parallel \hat z$, which yields the planar $g$-wave spin-splitting form.

Starting from the cubic parent space group $Pm\bar{3}m$, both the antiferroaxial distortion and the N\'{e}el order are associated with the $R$ point. Their condensation lowers the symmetry to the isotropy subgroup $I4/mcm$, leading to a planar $g$-wave altermagnetic spin splitting in the $xy(x^2-y^2)$ channel. The energetic barrier of the antiferroaxial distortion is quantified by the DFT total-energy profile as a function of $\theta$ [Fig.~\ref{fig:KMnF3}(c)]. By fitting this $E(\theta)$ curve, we extract an effective Landau free energy for the antiferroaxial order. Finally, we demonstrate that epitaxial strain provides an efficient tuning knob for the antiferroaxial order: the distortion can be switched on or suppressed by strain, and the magnitude obeys an approximately linear relation $\mathbf G^2 \propto \theta^2 \sim \varepsilon$ [Fig.~\ref{fig:KMnF3}(d)], consistent with the strain-coupling analysis in Sec.~\ref{sec:3}. When $\varepsilon_{zz} < 94.5\%$, the angle $\theta=0$. Consequently, the spin splitting vanishes, and the system effectively reduces to a $T\tau$-invariant antiferromagnet.

\textit{Computational details (KMnF$_3$):} We used $U=5$ eV on Mn, a $7\times 7\times 5$ k mesh, and a plane-wave cutoff energy of 550 eV.

\begin{figure*}[htbp]
\includegraphics[width=0.4\textwidth]{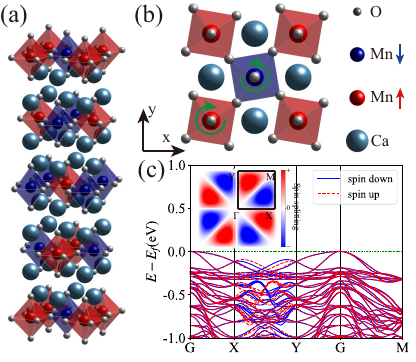}
\caption{\label{fig:Ca2MnO4}
The structure, antiferroaxial distortion and band structure of Ca$_2$MnO$_4$.
(a) Crystal structure of Ca$_2$MnO$_4$ (parent space group $I4/mmm$) with spin-up (red) and spin-down (blue) sublattices.
(b) The antiferroaxial distortion $G_z\sim P_2$ characterized by the counter-rotation of Mn-O$_6$ octahedra around the $z$-axis.
(c) Band structure exhibiting planar $g$-wave altermagnetic spin splitting.}
\end{figure*}

Ca$_2$MnO$_4$ crystallizes in the Ruddlesden-Popper perovskite structure~\cite{autretStructuralInvestigationCa2MnO42004}, and exhibits an analogous planar antiferroaxial rotation of MnO$_6$ octahedra around $\hat z$. This distortion lowers the symmetry from the parent space group $I4/mmm$ to isotropy subgroup $I4_1/acd$ and induces planar $g$-wave altermagnetic spin splitting. It exhibits an analogous planar antiferroaxial rotation of Mn-O$_6$ octahedra around $z$, as shown in Figs.~\ref{fig:Ca2MnO4}(a) and (b). Both antiferroaxial and antiferromagnetic phase transitions occur at the $P$ point with propagation vector $\mathbf{q}=(1/4,1/4,1/4)$, corresponding to the irreducible representations $G_z \sim P_2$ and $\mathbf{N} \sim P_1$, respectively. Upon condensation of $G_z$, the symmetry is lowered to the space group $I4_1/acd$. In this phase, the N\'{e}el vector transforms as $\mathbf{N} \sim \Gamma_2^+(A_{2g})$, which corresponds to a planar $g$-wave altermagnetism characterized by the $xy(x^2-y^2)$ symmetry [Table~\ref{table:altermagnet_valid_only}], as illustrated in Fig.~\ref{fig:Ca2MnO4}(c).

\textit{Computational details (Ca$_2$MnO$_4$):} We used $U=3$ eV on Mn and a $9\times 9\times 3$ k mesh (cutoff 550 eV).

\subsection{3D bulk $g$-wave altermagnets: MF$_3$ (M = Fe, Co, Ni) and LaCrO$_3$}
\label{subsec:FeF3_family}

\begin{figure*}[htbp]
\includegraphics[width=0.90\textwidth]{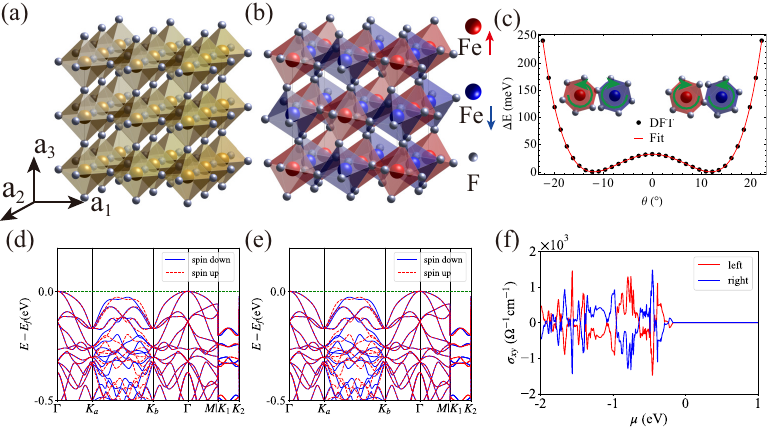}
\caption{\label{fig:FeF3}
Multiferroic switching in the bulk $g$-wave altermagnet FeF$_3$.
(a) The parent phase $Pm\bar{3}m$ with ideal octahedra. The antiferroaxial transition involves counter-rotations along the (111) direction (b).
(c) Total energy profile as a function of the rotation angle $\theta$, showing a double-well potential. The effective Landau free energy per unit cell is well fit to $\mathcal{F}_\mathrm{eff}(\theta)= 32.1-0.446\theta^2+1.45\times 10^{-3} \theta^4$,
where $\mathcal{F}_\mathrm{eff}$ is in meV per unit cell and $\theta$ is in degrees.
(d) Band structure for the equilibrium state $\theta=12^\circ$.
(e) Band structure for the reversed state $\theta=-12^\circ$, showing inverted spin splitting.
(f) Anomalous Hall conductivity (AHC) as a function of chemical potential $\mu$ for $\theta=\pm 12^\circ$, demonstrating the sign reversal of the transport signature.}
\end{figure*}

FeF$_3$ [Fig.~\ref{fig:FeF3}] and its isostructural counterparts (CoF$_3$, NiF$_3$) host a bulk antiferroaxial distortion of corner-sharing M-F$_6$ octahedra~\cite{leeWeakferromagnetismCoF3FeF32018,mattssonDensityFunctionalTheory2019}. The high-temperature parent phase is $Pm\bar{3}m$. At low temperatures, both antiferroaxial and antiferromagnetic instabilities occur at the $R$ point $\mathbf{q}=(\pi,\pi,\pi)$.
Symmetry analysis gives $\mathbf{G}\sim R_4^+$, $\mathbf{N}\sim R_1^+$, so that the bilinear combination transforms as $\mathbf{G}\mathbf{N}\sim R_4^+\otimes R_1^+ = \Gamma_4^+(T_{1g})$. Since the rank four magnetic multipole contains the $T_{1g}$ spatial channel, a symmetry-allowed trilinear coupling is present, stabilizing a bulk $g$-wave altermagnetic state.

The antiferroaxial magnitude is parameterized by the octahedral rotation angle $\theta$. The total-energy profile exhibits a double-well potential with degenerate minima at $\theta=\pm 12^\circ$ and a barrier of order $\Delta E\sim 30$ meV. The two antiferroaxial domains are related by a spin-group operation $[C_2||\boldsymbol{\tau}]$ with $\boldsymbol{\tau}=(0,0,1/2)$, enforcing a sign reversal of the nonrelativistic spin splitting upon $\theta\to-\theta$. With SOC included, the two domains are related by $\mathcal{T}\boldsymbol{\tau}$, implying that all time-reversal-odd response functions (e.g., anomalous Hall conductivity) must also reverse sign, as shown in Fig.~\ref{fig:FeF3}(d)

For FeF$_3$, the anomalous Hall conductivity was evaluated using a Wannier-interpolated tight-binding model and the standard Kubo formula~\cite{wangInitioCalculationAnomalous2006,yaoFirstPrinciplesCalculation2004},
\begin{equation}
\sigma_{xy}=\frac{e^2}{\hbar}\int_\mathrm{BZ}\frac{d^3k}{(2\pi)^3}\sum_{m\neq n}(f_{m\mathbf{k}}-f_{n\mathbf{k}})
\frac{\mathrm{Im}\!\left[\langle \psi_{n\mathbf{k}}|v_x|\psi_{m\mathbf{k}}\rangle
\langle \psi_{m\mathbf{k}}|v_y|\psi_{n\mathbf{k}}\rangle\right]}
{(E_{m\mathbf{k}}-E_{n\mathbf{k}})^2},
\end{equation}
where $v_\alpha=\partial H_{\mathbf{k}}/\partial k_\alpha$ ($\alpha=x,y$), $f_{n\mathbf{k}}$ is the Fermi-Dirac distribution, and $E_{n\mathbf{k}}$ denotes the band energy.

\textit{Computational details (FeF$_3$):} We used $U=5$ eV on Fe $3d$ orbitals and a plane-wave cutoff energy of 550 eV. The Brillouin-zone sampling was chosen consistently for a 3D bulk primitive cell.

\begin{figure*}[htbp]
\includegraphics[width=0.6\textwidth]{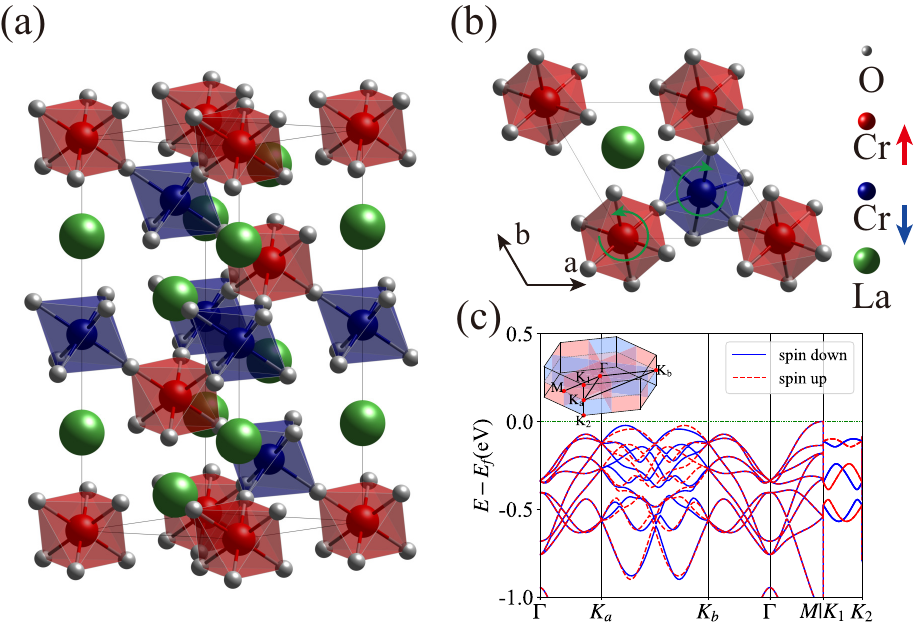}
\caption{\label{fig:LaCrO3}
The structure, antiferroaxial distortion and band structure of LaCrO$_3$.
(a) Crystal structure of LaCrO$_3$ (parent space group $Pm\bar{3}m$).
(b) The antiferroaxial distortion corresponding to the $R_4^+$ rotation mode of the Cr-O$_6$ octahedra (parent group $Pm\bar{3}m$).
(c) Band structure exhibiting bulk $g$-wave altermagnetic spin splitting.}
\end{figure*}

\begin{figure*}[htbp]
\includegraphics[width=0.6\textwidth]{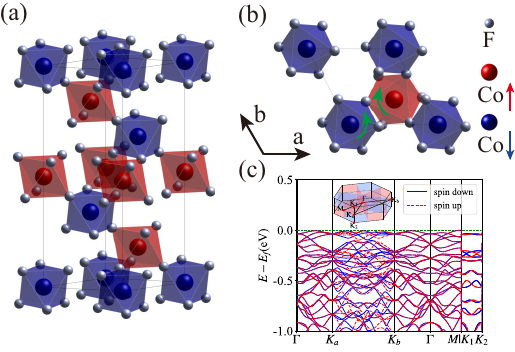}
\caption{\label{fig:CoF3}
The structure, antiferroaxial distortion and band structure of CoF$_3$.
(a) Crystal structure of CoF$_3$ (parent space group $Pm\bar{3}m$).
(b) The antiferroaxial distortion corresponding to the $R_4^+$ rotation mode of the Co-F$_6$ octahedra.
(c) Band structure exhibiting bulk $g$-wave altermagnetic spin splitting.}
\end{figure*}

LaCrO$_3$ [Fig.~\ref{fig:LaCrO3}]~\cite{zhouMagneticStructure$mathrmLaCrO_3$2011} and CoF$_3$ [Fig.~\ref{fig:CoF3}] shares the same $R_4^+$-type octahedral rotation mode and therefore realizes the same bulk $g$-wave antiferroaxial altermagnetism [Fig.~\ref{fig:LaCrO3} and Fig.~\ref{fig:CoF3}]. For LaCrO$_3$ and CoF$_3$, we used $U=3$ eV for Cr in LaCrO$_3$ and $U=6$ eV for Co in CoF$_3$, respectively (cutoff 550 eV).

\subsection{3D $d$-wave altermagnet: La$_2$NiO$_4$}
\label{subsec:La2NiO4}

\begin{figure*}[htbp]
\includegraphics[width=0.8\textwidth]{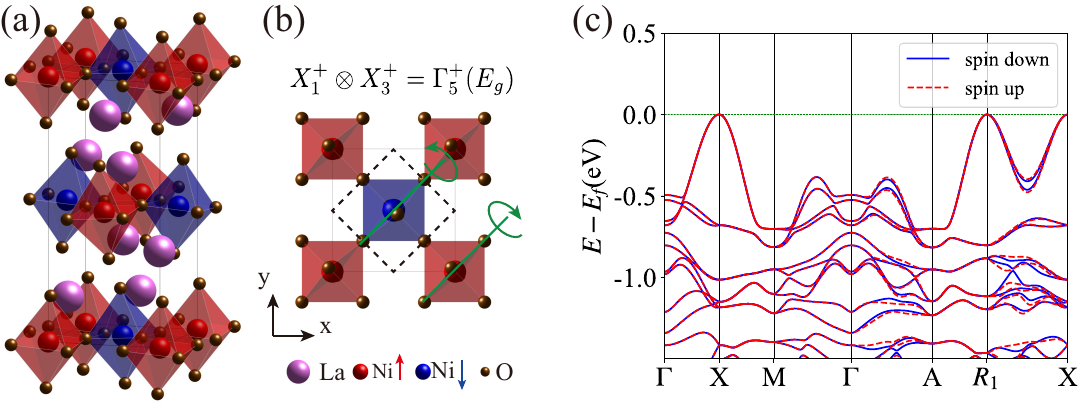}
\caption{\label{fig:La2NiO4}
The structure, antiferroaxial distortion and band structure of La$_2$NiO$_4$.
(a) Crystal structure of La$_2$NiO$_4$ (parent space group $I4/mmm$).
(b) Antiferroaxial distortion associated with the $X$-point mode $G_{110}\sim X_3^+$, involving rotations of NiO$_6$ octahedra about in-plane diagonal axes.
(c) Band structure exhibiting $d$-wave altermagnetic spin splitting in the ordered phase.}
\end{figure*}

La$_2$NiO$_4$ is a Ruddlesden-Popper compound with parent space group $I4/mmm$~\cite{rodriguez-carvajalNeutronDiffractionStudy1991}. Symmetry-mode analysis indicates concomitant antiferroaxial and antiferromagnetic instabilities at the $X$ point $\mathbf{q}=(\pi,\pi,0)$, yielding $G_{110}\sim X_3^+$ and $\mathbf{N}\sim X_1^+$. Their bilinear combination transforms as
\begin{equation}
G_{110}\mathbf{N}\sim X_3^+\otimes X_1^+ = \Gamma_5^+(E_g),
\end{equation}
which couples to the $E_g$ channel of the rank $\ell=2$ magnetic multipole, indicating a $d$-wave altermagnetic state. Upon condensation, $G_{110}$ lowers the crystallographic symmetry to $I4_2/ncm$, and the N\'{e}el order further reduces it to the space group $Pccn$ (point group $mmm$), in which the N\'{e}el order transforms as $\Gamma_3^+$, consistent with the $d_{xy}$-wave altermagnetic spin splitting shown in Fig.~\ref{fig:La2NiO4}(c).

\textit{Computational details:} We used $U=6$ eV for atom Ni and a $9\times 9\times 4$ k mesh (cutoff 550 eV).

\bibliography{ref}